\documentclass[
reprint,
superscriptaddress,
aps,
prd
]{revtex4}

\usepackage{babel,comment}
\usepackage{diagbox}
\usepackage{multirow}
\usepackage{soul}
\usepackage{amsmath,amsfonts,mathrsfs,amssymb}
\usepackage{latexsym}
\usepackage{graphicx}
\usepackage{dcolumn}
\usepackage{bm}
\usepackage{bbm}
\usepackage[T1]{fontenc} 
\usepackage{amsthm}
\usepackage{color}
\usepackage{slashed}
\usepackage{float}
\usepackage{esvect}
\usepackage{mathtools}
\usepackage{hyphenat}
\newcommand{\RNum}[1]{\uppercase\expandafter{\romannumeral #1\relax}}

\usepackage{cancel,soul,ulem}
\renewcommand{\sout}{\bgroup \color{red} \ULdepth=-.5ex \ULset}
\usepackage[pdfencoding=auto]{hyperref}

\begin{document}

\title{Thermal properties of the scalar glueballs from holography}

\author{Ruixiang Chen}
\email[]{chenruixiang22@mails.ucas.ac.cn}
\affiliation{School of Nuclear Science and Technology, University of Chinese Academy of Sciences, Beijing 100049, China}	

\author{Danning Li}
\email[]{lidanning@jnu.edu.cn}
\affiliation{Department of Physics and Siyuan Laboratory, Jinan University, Guangzhou 510632, P.R. China} 
	
\author{Mei Huang}
\email[]{huangmei@ucas.ac.cn}
\affiliation{School of Nuclear Science and Technology, University of Chinese Academy of Sciences, Beijing 101408, China}	

\begin{abstract}
 Based on a machine learning holographic QCD model, we construct a systematical framework to investigate the properties of the scalar glueballs continuously from zero temperature to finite temperature. By using both the quasi-normal frequencies and the spectral functions, we extract the pole masses, thermal widths, screening masses and dispersion relation of the scalar glueballs in hot medium. It is shown that the pole masses almost remain the vacuum values at temperatures far below the critical temperature $T_c$ , and then decrease with the increasing of temperature until a temperature lower than $T_c$. This result qualitatively agrees with earlier lattice simulations. While the pole masses increase monotonically above the critical temperature $T_c$, which agrees with recent lattice calculation. Meanwhile, it is shown that the thermal widths increase monotonically with temperature, which also agrees with the near $T_c$ lattice simulations. The screening mass exhibits a similar temperature-dependent behavior to the pole mass, while the dispersion relation increasingly deviates from the relativistic one as the temperature rises. It is interesting to note that we obtain the imaginary corrections in the thermal correlators, which contains both the temperal and spatial information and might be helpful for the four-dimensional calculations. Furthermore, by comparing the quasi-normal modes and the spectral functions, we note that it requires more careful analysis when applying the spectral functions in studying thermal hadrons from holography, since there could be other types of quasi-normal modes which are not related with bound states while they may contribute to the peaks of the spectral functions. 
\end{abstract}

\maketitle

\section{Introduction}

Glueball is the most direct prediction from the gluon self-interactions feature of Quantum Chromodynamics (QCD),  which attracts much attention from both theory and experiment. The glueball spectra has been investigated by using nonperturbative methods, e.g. lattice QCD calculations \cite{Morningstar:1999rf,Chen:2005mg,Gregory:2012hu}, QCD sum rules \cite{Latorre:1987wt,Huang:1998wj,Qiao:2014vva,Chen:2021cjr}, and recently by using holographic QCD method \cite{Colangelo:2007pt,DanningDHM13,FolcoCapossoli:2016fzj,Chen:2015zhh,Zhang:2021itx} as well as in Sakai-Sugimoto model \cite{Leutgeb:2019lqu,Brunner:2018wbv,Brunner:2015oga}. From experimental side, searching for glueballs has also made much progress \cite{Crede:2008vw}, and recently the evidence of C-odd odderons were announced by the D0 and TOTEM Collaborations\cite{D0:2020tig}, and the glueball-like particle X(2370) was observed at BESIII \cite{BESIII:2010gmv,Liu:2024anq}.

The investigation on glueballs, which are the only hadronic excitations at zero temperature in pure SU(3) Yang-Mills theory, is of essential meaning both in understanding the non-perturbative perspective of the strong interaction and in revealing the dynamics of the deconfinement phase transition. For the scalar glueballs, the lattice simulations \cite{MorningstarLat99,ChenYLat06,AthenodorouLat20,MeyerLat04} predict a ground state mass of around $1.475-1.730~\rm{GeV}$, and indicate a linear dependence of the mass square on the radial excitation quantum number $n$, i.e. $m_n^2\sim n$, which could be owing to the linear confinement.  At finite temperature, the breaking of Lorentz symmetry may lead to a splitting between the pole masses and the screening masses, which characterize temporal correlations and spatial correlations respectively.  For the temporal aspect, the early lattice simulations find decreasing pole masses of the lowest-lying scalar glueball with the increase of temperature before the deconfinement phase transition \cite{Ishii:2001zq,Ishii:2002ww,Meng:2009hh}. However, when including a constant contribution in the meson correlations, the recent lattice calculation \cite{Arikawa:2025kjx} observe a different behavior, with the pole masses staying unchanged below $T_c$ and increasing with temperature above $T_c$. Thus, it is still useful to conduct further research to examine whether the constant term originating from the deconfinement transition can have a significant impact in the low-temperature region. Moreover, compared with the temporal correlations, the spatial correlations of the scalar glueballs seems to be less studied. So, it is still interesting to investigate the full thermal properties of glueballs using other non-perturbative methods to obtain more detailed and valuable information.

As an important approach in dealing with the strong coupling problems in quantum field theories, the holographic method based on the anti-de Sitter/conformal field theory correspondence (AdS/CFT)  \cite{Maldacena:1997re,Gubser:1998bc,Witten:1998qj} turns out to be an effective and powerful method. A famous example is the prediction of the small value of shear viscosity to entropy density $\eta/s=1/(4\pi)$ \cite{Policastro:2001yc,Buchel:2003tz,Kovtun:2004de}, which matches the order used in fitting the data from the Relativistic Heavy Ion Collider (RHIC) \cite{Teaney:2000cw,Huovinen:2001cy,Hirano:2005xf,
Romatschke:2007mq}. A variety of models have been developed within this framework, including bottom-up models such as the hard-wall model \cite{Erlich:2005qh}, the soft-wall model \cite{Karch:2006pv}, V-QCD model \cite{Gursoy:2017wzz}, light-front holographic QCD \cite{Brodsky:15lfhQCD}, deformed AdS metric models and improved holographic QCD models \cite{Ghoroku:2005vt,Gherghetta:2009ac,Sui:2009xe}, which give good descriptions of hadronic physics, thermodynamics, transport properties of QCD matter.

The scalar glueball spectra of the vacuum from lattice simulations can be well described by modifying the metric or the dilaton field in dynamical holographic QCD which incorporates a gravity-dilaton system as the background, allowing for a more self-consistent description of the coupling between the dilaton field and the metric \cite{DanningDHM13,Zhang:2021itx}. It is interesting to see that by dynamically solving the dilaton system, the linear confinement in the glueball spectra can be well characterized. Besides the scalar glueball spectra, the higher spin pomeron or odderon spectra can also be described by introducing the tensor fields \cite{Boschi-Filho:2005xct,FolcoCapossoli:2013eao,FolcoCapossoli:2016uns,FolcoCapossoli:2016fzj}.  At finite temperature, Ref.\cite{Colangelo:2009ra} introduce the AdS-Schwarzchild metric in the quadratic dilaton model and read the thermal properties from the spectral functions. It gives a decreasing pole masses at very low temperatures. Meanwhile, Ref.\cite{Miranda:TAdSSW09} employed the same model but obtained the thermal properties through the corresponding quasi-normal frequencies, thereby capturing the behavior at higher temperatures. Due to the simplicity of the AdS-Schwarzschild metric and its lack of information about the phase transition, the behavior at finite temperature and its relation with the critical temperature still require further investigation. On the other hand, in order to get more understanding on the thermal properties, it is still quite necessary to do the investigation in a holographic model which has been well established and tested in many other aspects, at least the thermodynamics. 

In this work, to compare with the lattice simulations, we will focus on the scalar glueballs in the pure gluon system. So its dual description would be scalar perturbative above the dual quenched gravity model. When extending this model to finite temperature and finite density systems, additional degrees of freedom are introduced to form the Einstein-Maxwell-Dilaton system as the background \cite{Gursoy:2008bu,Gubser:2008ny,DeWolfe:2010he,He:2013qq,Yang:2014bqa,Yang:2015aia,Dudal:2017max,Dudal:2018ztm,Fang:2015ytf,Li:2011hp,Li:2022erd,Critelli:2017oub,Grefa:2021qvt,Arefeva:2020vae,Chen:2018vty,Chen:2020ath}. This approach can quantitatively describe the strongly coupled Quark-Gluon Plasma (QGP) under extreme conditions. When using this model to solve problems, it is necessary to provide an ansatz by specifying the form of one of the following: the dilaton field, the warp factor in the metric, or the potential function of the dilaton field. Thus there are two predominant methods to obtain solutions of the EMD model. The first involves inserting the form of the dilaton potential and numerically constructing a family of five-dimensional black holes \cite{Gubser:2008ny,DeWolfe:2010he,Critelli:2017oub,Grefa:2021qvt,He:2022amv}. The second method, known as the potential reconstruction method \cite{Li:2011hp,He:2013qq,Yang:2014bqa,Yang:2015aia,Dudal:2017max,Fang:2015ytf,Chen:2018vty,Chen:2020ath,Zhou:2020ssi}, involves inputting the dilaton or a metric function to determine the dilaton potential. Although this approach results in a temperature-dependent dilaton potential, the model can nevertheless capture many QCD properties through analytical solutions. However, an ansatz that performs well for solving one type of problem may prove inadequate when applied to another type of problem. For example, a quadratic dilaton field can yield very accurate results when solving for the hadronic spectra at zero temperature, but when extending the system to finite temperature by introducing a black hole, this choice imposes a relatively high minimum temperature limit below by which it is hard to obtain the temperature dependent properties of hadrons.

Furthermore, with the advancement of deep learning, related applications have emerged in the field of holographic QCD recently \cite{Hashimoto:2022eij,PhysRevD.109.L051902}. This technique allows for more exploration of the forms of the introduced fields. In Ref.\cite{PhysRevD.109.L051902}, the authors constructed a neural network to learn the equation of state results and baryon number susceptibility at finite temperature and density, thereby obtaining a specific form of the warp factor. Since this model has been carefully examined in describing thermodynamic quantities, we will extend it to study the thermal properties of the scalar glueballs. 

In this paper, we will use a modified dynamical holographic models with the aforementioned warp factor to calculate the pole mass, thermal width, screening mass, and dispersion relation of the scalar glueball at finite temperature. In Section \ref{sec:Holography Model}, we will introduce the setup of the holographic model and present the black hole solutions derived from the equations of motion after inputting the specific form of the warp factor into the background action. In Section \ref{sec:Spectra of glueballs}, we will construct a holographic model describing the scalar glueball and use it to study the glueball's properties at both zero and finite temperatures. At zero temperature, we will calculate the glueball spectrum and fix the model. At finite temperature, we employ two complementary methods—calculating the spectral function and determining the complex frequencies of quasi-normal modes (QNMs)—to extract the mass and width of glueballs. Additionally, by analyzing the poles of the two-point correlation function, we obtain the screening mass and dispersion relations at various temperatures. Finally, in Section \ref{sec:Discussion and Conclusion}, we will provide a summary and discuss the results.

\section{Holography Model}
\label{sec:Holography Model}

To compare with the lattice simulations, we will focus on the pure gluon system and try to consider the corresponding excitations in a dual thermal background. The total action of the five-dimensional holographic QCD model including glueball excitations can generally be taken as

\begin{equation}
    S_{\text{total}}^s=S_b^s+S_g^s,     
\end{equation}

\noindent where $S_b^s$ is the action for the background in the string frame, and $S_g^s$ is the action describing glueballs in the string frame. Since the glueballs are excitations on the thermodynamic background and will be considered as a perturbation, here, $S_g^s$ will not affect the thermodynamics, i.e., taking the probe limit. Therefore, in this section, we will provide a detailed description of the background action and its solutions, while the action describing glueballs will be left in the next section.

To describe QCD matter at finite temperature, a simple choice is to consider the gravity-dilaton system. The gravity-dilaton action $S_b^s$ for the background includes a gravity field $g_{M N}^s$ and a neutral dilatonic scalar field $\Phi_s$, which in string frame is given by

\begin{equation}
    S_b^s=\frac{1}{16 \pi G_5} \int d^5 x \sqrt{-g^s} e^{-2 \Phi_s}\left[R_s+4 \partial_M \Phi_s \partial^M \Phi_s-V_s\left(\Phi_s\right)\right],
\end{equation}

\noindent where $V\left(\Phi\right)$ is the potential of the dilaton field and $G_5$ is the Newton constant in five dimensions. Generally, one can choose a specfic form of the dilaton potential $V\left(\Phi\right)$ and solve the system. In this work, we will take the potential reconstruction approach, in which the explicit forms of the dilaton potential can be solved consistently from the equations of motion (EoM) after assuming the profiles of the dilaton field or the warp factor of metric.

To facilitate the derivation of the equations of motion later on, we transform the action from string frame to Einstein frame using the following transformations:

\begin{equation}
    \Phi_s=\sqrt{\frac{3}{8}} \Phi, \quad g_{M N}^s=g_{M N} e^{\sqrt{\frac{2}{3}} \Phi},  \quad V_s\left(\Phi_s\right)=e^{-\sqrt{\frac{2}{3}} \Phi} V(\Phi).
\end{equation}

\noindent The background actions in Einstein frame becomes

\begin{equation}
    S_b=\frac{1}{16 \pi G_5} \int d^5 x \sqrt{-g}\left[R-\frac{1}{2} \partial_M \Phi \partial^M \Phi-V(\Phi)\right].
\end{equation}

The EoMs derived from the action are, respectively,

\begin{equation}
\begin{gathered}
    R_{M N}-\frac{1}{2} g_{M N} R-T_{M N}=0,\\
    \partial_M\left[\sqrt{-g} \partial^M \Phi\right]-\sqrt{-g}\frac{\partial V}{\partial \Phi}=0,
\end{gathered}
\end{equation}

\noindent with

\begin{equation}
\begin{aligned}
T_{M N}= & \frac{1}{2}\left(\partial_M \Phi \partial_M \Phi-\frac{1}{2} g_{M N}(\partial \Phi)^2-g_{M N} V(\Phi)\right).
\end{aligned}
\end{equation}

We take the following ansatz of metric

\begin{equation}
    d s^2=\frac{L^2 e^{2 A(z)}}{z^2}\left[-f(z) d t^2+\frac{d z^2}{f(z)}+d \vec{x}^2\right],
\end{equation}

\noindent where $z$ is the 5th-dimensional holographic coordinate and the AdS radius $L$ of $\rm AdS_5$ space is set to be one (in fact, when one considers a flat boundary, $L$ could be absorbed in other dimensional parameters everywhere).  The bulk geometry of this form is a black hole solution with an event horizon $z_h$ where the blackening factor $f(z)$ vanishes. Using the above ansatz of the metric, the equations of motion and constraints for the background fields can be obtained as

\begin{equation}
\begin{gathered}
    \Phi^{\prime \prime}+\Phi^{\prime}\left(-\frac{3}{z}+\frac{f^{\prime}}{f}+3 A^{\prime}\right)-\frac{L^2 e^{2 A}}{z^2 f} \frac{\partial V}{\partial \Phi}=0,
\end{gathered}
\end{equation}

\begin{equation}
    f^{\prime \prime}+f^{\prime}\left(-\frac{3}{z}+3 A^{\prime}\right)=0,
\end{equation}

\begin{equation}
\begin{aligned}
    A^{\prime \prime} & +\frac{f^{\prime \prime}}{6 f}+A^{\prime}\left(-\frac{6}{z}+\frac{3 f^{\prime}}{2 f}\right)-\frac{1}{z}\left(-\frac{4}{z}+\frac{3 f^{\prime}}{2 f}\right)+3 A^{\prime 2} +\frac{L^2 e^{2 A} V}{3 z^2 f}=0,
\end{aligned}
\end{equation}

\begin{equation}
    A^{\prime \prime}-A^{\prime}\left(-\frac{2}{z}+A^{\prime}\right)+\frac{\Phi^{\prime 2}}{6}=0,
\end{equation}

\noindent where the first equation comes from the equation of motion for the dilaton field, and the next three equations correspond to the z-direction part, the temporal part, and the spatial part of Einstein's equations respectively. Only three of the above four equations are independent.

To solve those equations, we need to specify the boundary conditions first. For the blackening factor near the horizon, one should impose

\begin{equation}
    f\left(z_h\right)=0.
\end{equation}

Near the boundary $z \rightarrow 0$, we require the metric in string frame to be asymptotic to $\rm AdS_5$. The boundary conditions are

\begin{equation}
    A(0)=-\sqrt{\frac{1}{6}} \Phi(0)=0, \quad f(0)=1.
\end{equation}

Since there are four unknown functions in the three independent equations, we need to specify the explicit form of one of them in order to solve for the others. When we choose $A(z)$ as the input function, the EoMs can be analytically solved as

\begin{equation}
\begin{aligned}
    \Phi^{\prime}(z) & =\sqrt{-6\left(A^{\prime \prime}-A^{\prime 2}+\frac{2}{z} A^{\prime}\right)}, \\
    f(z) & =1-\frac{\int_0^z y^3 e^{-3 A} d y }{\int_0^{z_h} y^3 e^{-3 A} d y }, \\
    V(z) & =-3 z^2 f e^{-2 A}\left[A^{\prime \prime}+3 A^{\prime 2}+\left(\frac{3 f^{\prime}}{2 f}-\frac{6}{z}\right) A^{\prime}-\frac{1}{z}\left(\frac{3 f^{\prime}}{2 f}-\frac{4}{z}\right)+\frac{f^{\prime \prime}}{6 f}\right].
\end{aligned}
\label{eq:background}
\end{equation}

To obtain a solution of the system, we will use the following ansatz:

\begin{equation}\label{eqAz}
    A(z)= d \ln(a z^2 + 1) + d \ln(b z^4 + 1),
\end{equation}

\noindent which is taken from Ref.\cite{PhysRevD.109.L051902},
with 4 undetermined parameters included $G_5$ which will be fixed by comparing the result of equation of state(EoS) with lattice results under the help of machine learning. The result is $a = 0, b = 0.072, d = -0.584, G_5=1.326$, and critical temperature $T_c=0.265$ GeV for pure gluon case. Here, though the first term in Eq.\eqref{eqAz} is irrelevant when $a=0$, we still keep it since it could be turned on to describe systems with flavors. 

With the above warp factor, the expression for $\Phi(z)$ can be obtained as

\begin{multline}
    \Phi(z)=\frac{\sqrt{438}}{125} \left(
    4 \sqrt{13} \, \mathrm{ArcTan} \left[\frac{125 \sqrt{417} + 9 \sqrt{417} z^4 + 3 z^2 \sqrt{78125 + 3753 z^4}}{500 \sqrt{13}}\right] + \right.\\
    \left.\sqrt{417} \, \mathrm{ArcTanh} \left[\frac{3 z^2}{\sqrt{\frac{78125}{417} + 9 z^4}}\right]\right) - \frac{4}{125} \sqrt{5694} \, \mathrm{ArcTan} \left[\frac{\sqrt{\frac{417}{13}}}{4}\right].
\end{multline}

\noindent It should be noted that the coefficients in the expression depend on the specific values of $b$ and $d$. It exhibits the following asymptotic behavior in the infrared and ultraviolet regions:

\begin{equation}
    \Phi \left( z\rightarrow 0 \right) =\frac{3}{25} \sqrt{\frac{438}{5}} z^2 - \frac{7497 \sqrt{\frac{219}{10}}}{1953125} z^6 + \mathcal{O}(z^7),\Phi \left( z\rightarrow \infty \right) =constant+\frac{6\sqrt{20294}}{125}\ln\left( z \right).
\end{equation}

After solving the solution, the temperature can be obtained from

\begin{equation}
    T=\frac{\left| f^\prime\left( z_h \right) \right|}{4\pi},
\end{equation}
and the corresponding thermodynamic quantities/properties could be studied.

\begin{figure}
\centering
\includegraphics[width=0.55\textwidth]{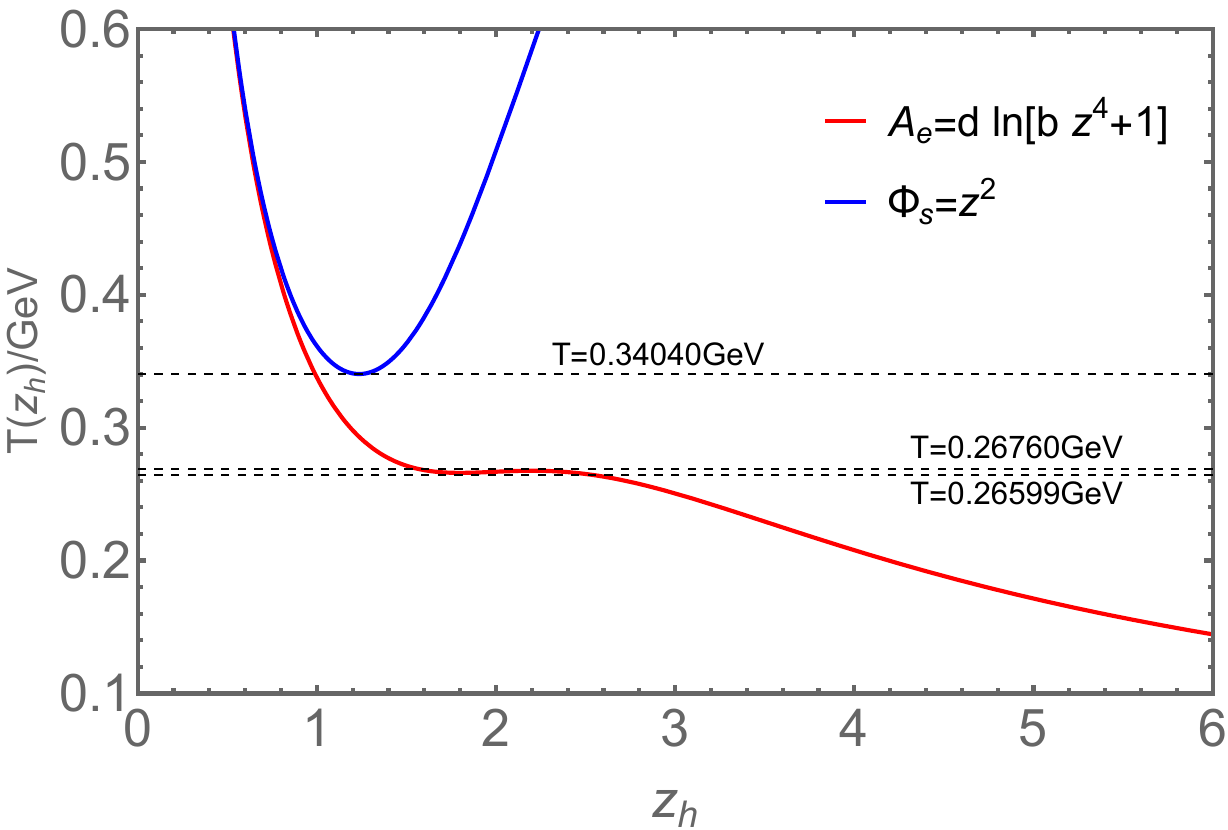}
\caption{\label{fig:T-zh/0} The temperature $T$ as a function of the black-hole horizon $z_h$. The red line represents the results obtained using $A(z)=d\, ln(az^2+1)+d\, ln(bz^4+1)$. In the range of [1.78, 2.20], the function exhibits a slight monotonic increase. The blue line represents the results obtained using $\Phi=kz^2$ while $A(z)$ is obtained by solving the equation of motion. There exists a minimum temperature $T_{min} \approx 0.34GeV$. }
\end{figure}

Using Eq.\eqref{eq:background}, one obtains the results of the relationship between temperature and the horizon which are shown in Fig.\ref{fig:T-zh/0}. It can be seen that, except for a small region $1.78 \rm{GeV}^{-1}\leq z\leq 2.20 \rm{GeV}^{-1}$ where the function increases with temperature, it decreases monotonically in other areas. This monotonically increasing region is near $T=T_c=0.265 \rm{GeV}$, where a phase transition happens as shown in \cite{PhysRevD.109.L051902}. Besides, the thermodynamic quantities of pure gluon system from such a simple model agree very well with the results from lattice simulation (for the details, please refer to Ref. \cite{PhysRevD.109.L051902}). It can also be proven that when $z_h$ is sufficiently large, the behavior of $T$ is similar to $\frac{1}{z_h}$, and it approaches 0 at infinity. For comparison, we also show in the figure the result obtained by introducing $\Phi(z)=z^{2}$ into the equation of motion. In this case, the system's temperature exhibits a minimum value. When $T<T_{min}$, the thermal gas solution is considered to be dominant\cite{JHEP05(2009)033}, where $A(z)$ and $\Phi(z)$ remain consistent with the black hole solution presented earlier but $f(z)=1$. This configuration still satisfies the equations of motion. So it inherits all the non-perturbative features(confinement, spectrum, values of condensates, etc.). In addition, in the black hole solution, for the region where the temperature increases with $z_h$, the entropy decreases as the temperature increases, which makes the system unstable and therefore this region should be avoided.

\begin{figure}
\centering
\includegraphics[width=0.6\textwidth]{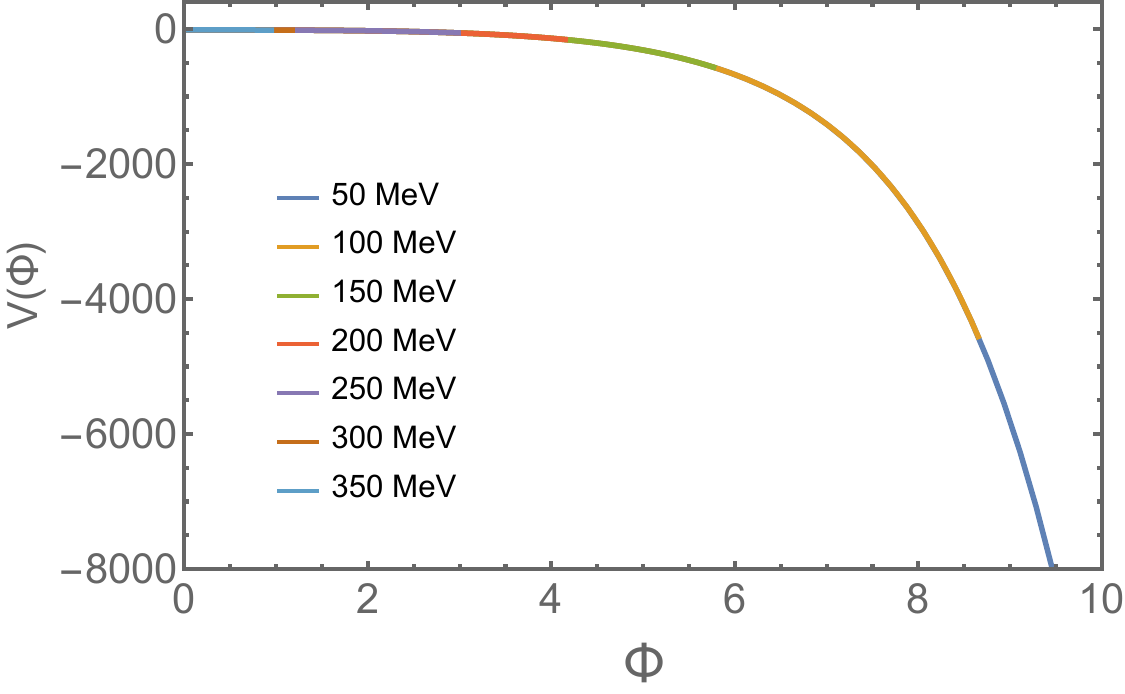}
\caption{\label{fig:V-phi} The dilaton potential $V(\Phi)$ at different temperature. }
\end{figure}

Generally, because we input $A_e$, whose form is independent of temperature, we can derive $V(\Phi)$ with
temperature dependent $V(\Phi,T)$. However, as shown in the Fig.\ref{fig:V-phi}, the difference in V at different temperatures are very small, with relative deviations less than $10^{-2}$. It is also noteworthy that Ref.\cite{PhysRevD.109.L051902} shows the dilaton potential in this model closely resembles the potential in the DGR model\cite{DeWolfe:2010he}.

\section{Spectra of glueballs}
\label{sec:Spectra of glueballs}

In this section, we discuss the spectra of  scalar glueballs at zero temperature and finite temperature. The action of the related glueball fields will be introduced and the glueballs will be treated as excitations above the background. At zero temperature, we obtain the glueball spectrum by solving the eigenvalue problem of the equations of motion. At finite temperature, we study the properties of the glueball spectrum by combining the calculation of the spectral function and the quasi-normal mode (QNM) frequencies.

In the string frame, the action describing scalar glueballs can be written as

\begin{equation}
    S_g^s=-c_g\int d^5 x \sqrt{-g_s} h(\Phi_s)\left(\frac{1}{2} g_s^{MN} \partial_{M} S \partial_{N} S+\frac{1}{2} M_5^2 S^2\right),
\label{eq: Glueball Action}
\end{equation}

\noindent where s denotes the string frame, $c_g$ describes the coupling strength of glueballs part in the whole theory\cite{ZhangLin:22glueball}. The scalar fields S (we will always use $S_{5D}$ to represent the five-dimensional action), is 5-dimensional fields that are dual to scalar glueball operator. 
The  5-dimensional mass $M^2_5$ is given by the $\text{AdS}_5/\text{CFT}_4$ correspondence dictionary. Based on the AdS/CFT dictionary, the conformal dimension $\Delta$ of a q-form operator at the ultraviolet (UV) boundary is related to the 5-dimensional mass square $M^2_5$ of its dual field in the bulk as follows:

\begin{equation}
    M_{5}^{2}=\left( \varDelta -q \right) \left( \varDelta +q-4 \right).
\end{equation}

\noindent For the scalar glueball, the dual operator is the dimension-4 scalar gluon operator $\text{Tr}(G^2)$ with $\Delta=4, q=0$. So $M^2_5$ can be easily obtained as $0$ and only the kinetic term is kept in Eq. \eqref{eq: Glueball Action}.

In the original soft-wall model, the coupling function $h(\Phi_s)$ is chosen as $h(\Phi_s)=e^{-\Phi_s}$.  It is not difficult to accept such a coupling from the point of view of open string coupling. However, since the scale of the string is far from the scale of phenomenological interests, it is also acceptable to apply other forms of this factor by fitting the experimental observables. As we will see later, with the former case $h=e^{-\Phi_s}$, the model fails to produce bound states for glueballs. Therefore we will take the latter point of view and fix the factor $h(\Phi_s)$ by the vacuum spectra of the scalar glueball from lattice simulations. It is noticed that a recent machine learning work \cite{Chen:2025kqb} showed a modified metric can combine both glueball spectra at zero temperature and equation of state. However, this modified model with $h=e^{-\Phi_s}$ still cannot investigate the system continuously from zero to high temperature. 

Furthermore, to reduce the dependence of the final conclusion on the form of this factor, we will take three kinds of the form of $h(\Phi_s)$. Since $\Phi(z)$ is monotonical, it is possible to take $h$ as a function of $z$ and obtain the numerical form of it as a function of $\Phi$.:
\begin{align}
    h(z)&=e^{-kz^2}\,(Choice I),\\
    h(z)&=e^{-\Phi_{s}-lz^{4}}\,(Choice II),\\
    -ln[h(z)]&=3A_{s}(z)-jz^2+3ln \left[ \frac{2^{\frac{3}{8}} \times 3^{\frac{1}{8}} \Gamma\left( \frac{5}{4} \right) I_{\frac{1}{4}}\left( \frac{\sqrt{\frac{8}{3}}j z^2}{\sqrt{6}} \right)}{\sqrt{\frac{8}{3}} j^{\frac{1}{4}} \sqrt{z}}\right]\,(Choice III),
\end{align}

\noindent where $j$, $k$, and $l$ are parameters, $\Gamma(z)$ is the Euler gamma function, and $I_{n}(z)$ is the modified Bessel function of the first kind.  For later convenience, we will call Choice I/II/III as MDHM ( modified dynamical HQCD model ) I/II/III respectively.

We now provide some explanation for the complicated form of Choice III. In Ref.\cite{DanningDHM13}, by substituting $\Phi_s(z)=jz^2$ into the equations of motion, one obtains

\begin{equation}
    A(z)= ln \left[ \frac{2^{\frac{3}{8}} \times 3^{\frac{1}{8}} \Gamma\left( \frac{5}{4} \right) I_{\frac{1}{4}}\left( \frac{\sqrt{\frac{8}{3}}j z^2}{\sqrt{6}} \right)}{\sqrt{\frac{8}{3}} j^{\frac{1}{4}} \sqrt{z}}\right] 
\end{equation}

\noindent and can derive a glueball spectrum that fits well with the lattice results by setting $j=1\text{GeV}^2$. If we denote the two equations above as $\Phi_{s,quadratic}$ and $A_{quadratic}$ respectively (note that $A_{quadratic}$ is in the Einstein frame), and then set $-ln(h(z))=3A_s(z)-3A_{s,quadratic}+\Phi_{s,quadratic}$, according to Eq.\eqref{eq:Veff}, we can see that this leads to an effective potential for the glueball equation of motion in our model that is exactly the same as that obtained in the model used in Ref.\cite{DanningDHM13}, which implies that it can produce an identical glueball spectrum. The purpose of doing this is to demonstrate the diversity of possible forms of $h(\Phi)$ that can lead to glueball spectra in good agreement with lattice results. Moreover, if the three chosen forms yield similar results when studying the properties of glueballs at finite temperature (as will be shown in the following sections), this would strengthen the model-independence of the related conclusions.

In the calculation process, this is equivalent to keeping the $e^{-\phi_s(z)}\mathcal{L}_{g}^{s}$ coupling form while setting $\phi_s=-ln[h(z)]$. Therefore, to keep the formulas in this paper as consistent as possible with those in other works using holographic methods, the $\phi_{s}$ that appears later is defined in this way, rather than simply referring to the dilaton field in the string frame $\Phi_s$.

After the above clarification, we can derive the EoM from the action $S_g^s$:
\begin{equation}
\partial _z\left( \sqrt{-g_s}e^{-\phi _s}g_{s}^{zz}\partial _zS \right) +\sqrt{-g_s}e^{-\phi _s}g_{s}^{\mu \nu}\partial _{\mu}\partial _{\nu}S=0,
\label{EoM}
\end{equation}
and in the momentum space, it satisfies:
\begin{equation}
\partial _z\left(\frac{e^{3A_s-\phi _s}}{z^3}f\left( z \right) \partial _zS\left( p,z \right) \right) +\frac{e^{3A_s-\phi _s}}{z^3}\left( \frac{\omega ^2}{f\left( z \right)}-\boldsymbol{p}^2 \right) S\left( p,z \right) =0.
\label{EoMp}
\end{equation}.

\subsection{Glueball at zero Temperature}
\label{subsec Glueball at zero Temperature}

\begin{figure}[t]
\centering
\includegraphics[width=0.5\textwidth]{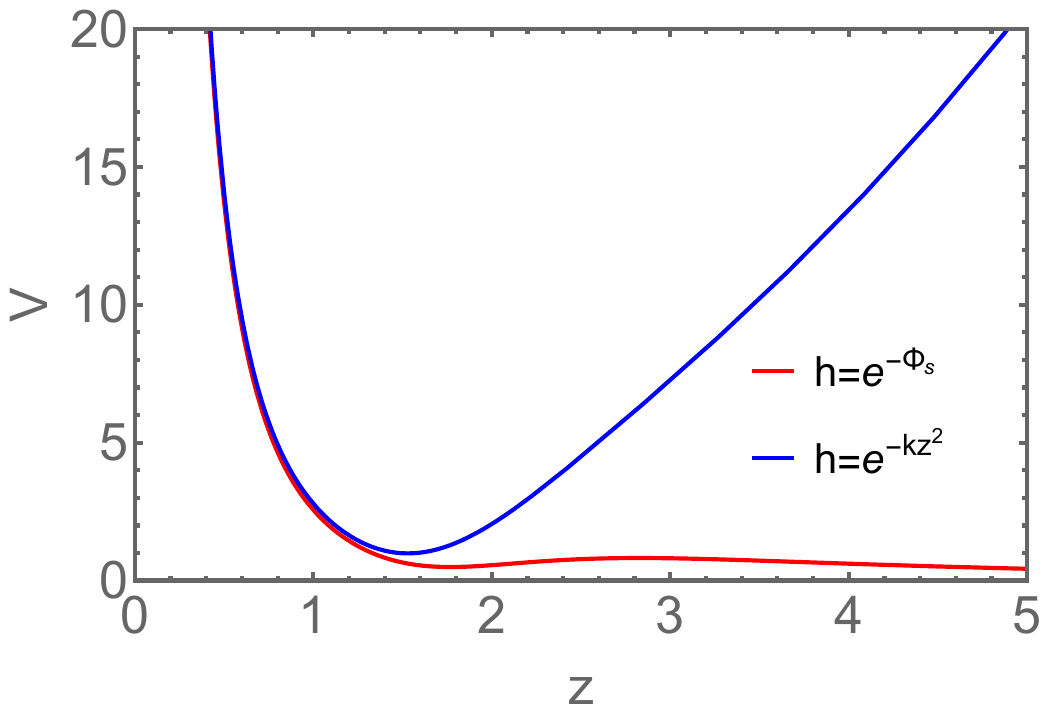}
\caption{\label{fig:Veff0} The effective potentials of scalar glueballs.}
\end{figure}

When considering the spectra at zero temperature, we have $f\left( z \right) =1$ and $z\in \left( 0,+\infty \right) $, and via the  substitution
\begin{equation}
S_n\rightarrow z^{\frac{3}{2}}e^{-\frac{1}{2}\left(3A_s-\phi_s \right)}S_n ,
\end{equation}
the EoM can be brought into Schrödinger-like equation:
\begin{equation}
 -\partial_{z}^{2}S_n+V_sS_n=m_{n}^{2}S_n,
\end{equation}
with the 5-dimensional effective Schrödinger potential

\begin{equation}
V_s=\frac{3A_{s}^{''}+\frac{3}{z^2}-\phi _{s}^{''}}{2}+\frac{\left( 3A_{s}^{\prime}-\frac{3}{z}-\phi _{s}^{\prime} \right) ^2}{4}.
\label{eq:Veff}
\end{equation}

\begin{figure}
    \centering
    \includegraphics[width=0.5\linewidth]{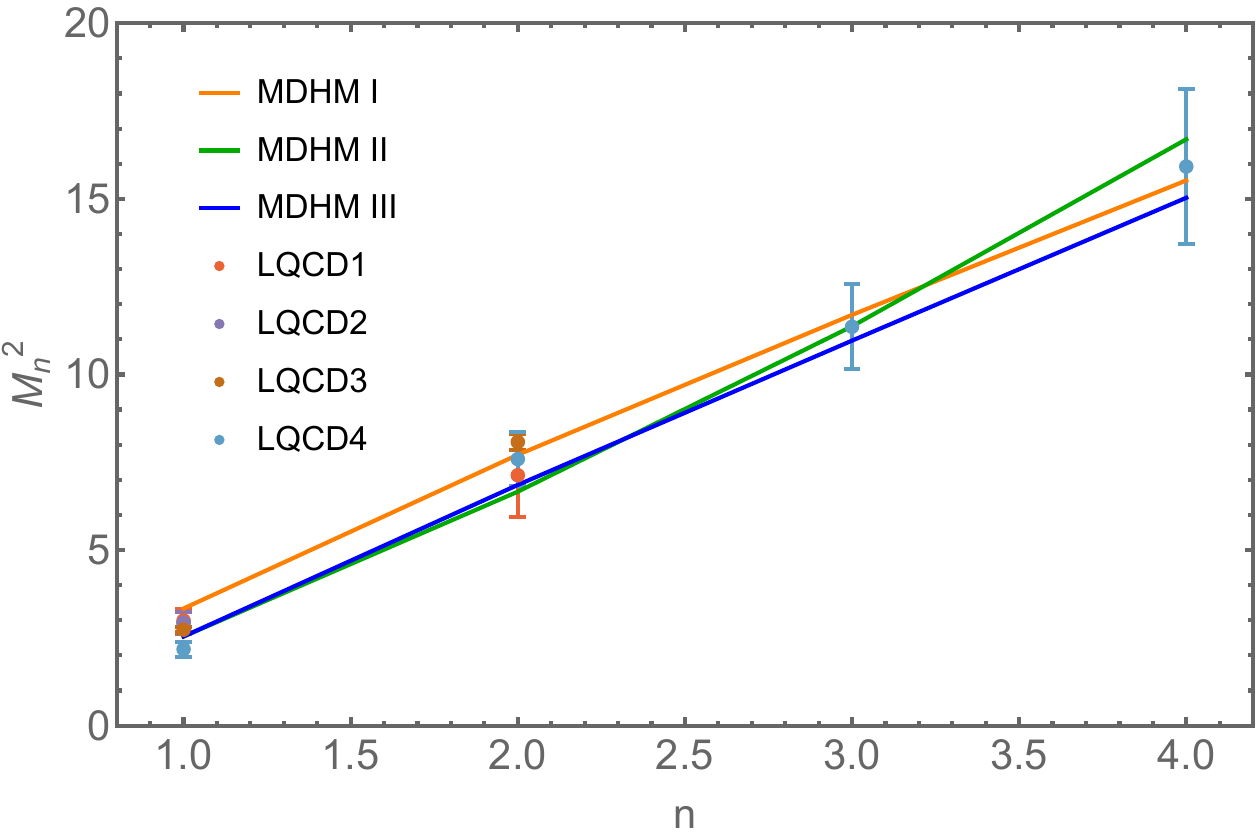}
    \caption{Comparison between the squared masses of scalar glueball excitations obtained from different choices and their dependence on the excitation number with the lattice results. The lattice results are also taken from \cite{MorningstarLat99,ChenYLat06,AthenodorouLat20,MeyerLat04}.}
    \label{fig:M2-n}
\end{figure}

\noindent As introduced in Ref.\cite{PhysRevD.74.015005}, $m_n^2=\omega^2-\boldsymbol{p}^2$ in this equation is the mass of the glueball, which can be obtained by solving for the eigenvalues of this Schrödinger equation. We show the shapes of the effective potentials for the cases $h=e^{-\Phi_s}$ and $h=e^{-kz^2}$ with $k=(0.95\rm{ GeV})^2$ respectively in Fig.\ref{fig:Veff0}. As shown in the figure, for the conventional choice($h=e^{-\Phi_s}$), although the effective potential has a very shallow well, it approaches zero as $z\rightarrow\infty$, indicating that bound states cannot be formed. Therefore, we attempt to overcome this issue by modifying the model. After adjusting the coupling form between the dilaton field and the glueball action, the resulting change in the effective potential enables the formation of glueball bound states. Taking the lattice results in \cite{MorningstarLat99,ChenYLat06,AthenodorouLat20,MeyerLat04} (shown in Tab.\ref{tab:glueball mass}), we fit the parameters left in the three choices as $k=0.95^{2}\,GeV^{2}$(I), $l=0.0354\,GeV^4$(II) and $j=1\,GeV^2$(III) and work out the glueball masses, as shown in the Tab.\ref{tab:glueball mass} and Fig.\ref{fig:M2-n}. It can be seen that a wide range of choices can yield results that agree well with the lattice data. However, to ensure regge trajectory that the squared glueball mass $M_{n}^{2}$ exhibits a linear relationship with the excitation number $n$ over a wide range, the behavior of $h$ needs to be similar to that of $e^{-kz^2}$ at large $z$.

\begin{table}
\renewcommand{\arraystretch}{1.5}
\setlength{\tabcolsep}{2mm}
\begin{tabular}{c|c|cccc}
\hline
 & n & 1 & 2 & 3 & 4 \\ \hline
MDHM I &  & 1.826 & 2.777 & 3.420 & 3.940 \\
MDHM II &  & 1.593 & 2.581 & 3.372 & 4.086 \\
MDHM III &  & 1.593 & 2.618 & 3.311 & 3.877 \\
LQCD1 & M /GeV & 1.710(50)(80) & ... & ... & ... \\
LQCD2 &  & 1.730(50)(80) & 2.670(180)(130) & ... & ... \\ 
LQCD3 &  & 1.653(26) & 2.842(40) & ... & ... \\
LQCD4 &  & 1.475(30)(65) & 2.755(70)(120) & 3.370(100)(150) & 3.990(210)(180)\\
\hline
\end{tabular}
\caption{\label{tab:glueball mass}The scalar glueball masses obtained using three different choices of $h(z)$ are presented(MDHM I-III). The parameters are set as $k=0.95^{2}\,GeV^{2}$(I), $l=0.0354\,GeV^4$(II) and $j=1\,GeV^2$(III), respectively. The lattice QCD results are taken from Ref.\cite{ChenYLat06,MorningstarLat99,AthenodorouLat20,MeyerLat04} respectively.}
\end{table}

It is also worth noting that the glueball action part of this model is quite similar to that of the soft-wall model, although the latter is set in an $\text{AdS}_5$ spacetime. The soft-wall model, however, struggles to simultaneously describe both the ground state mass and the Regge slope accurately. In contrast, by using the improved metric in our model, the results are significantly improved.

\subsection{Glueball at finite Temperature}

At finite temperature, due to the breaking of Lorentz symmetry, the original definition of mass is no longer appropriate. Instead, two distinct concepts are used: the pole mass and the screening mass, which describe the correlations of the two-point Green function in the temporal and spatial directions, respectively. The pole mass is considered the physically observable mass of a thermal hadron, while the screening mass characterizes the length scale of strong interactions within the thermal medium, i.e., $G\left( \boldsymbol{x} \right) \sim e^{-m_{scr}\left| \boldsymbol{x} \right|}/\left| \boldsymbol{x} \right|$ when $\left| \boldsymbol{x} \right|\gg m_{scr}^{-1}$. Moreover, due to interactions with the surrounding thermal medium, even originally stable hadrons may become unstable and acquire a thermal width.

On the other hand, these quantities can be related to the pole of the two-point correlation function in momentum space: the pole mass and thermal width correspond to the real and imaginary parts, respectively, of the pole in the retarded two-point correlation function, while the screening mass is given by the pole in the spatial correlation function.

In this section, we use the holographic method to calculate the correlation functions, and thereby obtain the pole mass, screening mass, and thermal width of the scalar glueball at finite temperature, as well as their temperature-dependent behavior.

In detail, the partitial function can be obtained from the equivalence of the two sides
\begin{equation}
    \left< e^{i\,\,\int{d^4xX_0\left( x \right) O_G\left( x \right)}} \right> _{CFT}=e^{iS_{5D}\left[ X\left( x,z \right) \right]},
\label{eq:GKB}
\end{equation}
where $S_{5D}\left[ X\left( x,z \right) \right]$ is the on-shell action of the bulk field $X(x,z)$ dual to $O_G\left( x \right)$, with $X_0(x)$ as a source term.  Therefore the two-point correlation function can be derived by differentiating the right-hand side of Eq.\eqref{eq:GKB} with respect to the source twice.

By combining Eq.\eqref{eq: Glueball Action} and Eq.\eqref{EoM}, we obtain the onshell 5-dimension action:

\begin{equation}
    S_{5D}^{on-shell}=\int{d^4x\frac{c_g}{2}\sqrt{g_s}e^{-\phi _s\left( z \right)}g^{zz}_{s}S\left( x,z \right) \partial _zS\left( x,z \right)}\mid_{z=0}.
\end{equation}

\noindent Then, the two-point correlation function is

\begin{equation}
\begin{aligned}
    G\left( \omega,\boldsymbol{p} \right) &=-i\int d^{4}x \frac{1}{i}\frac{\delta}{\delta S\left( x, 0 \right)}\frac{\delta S_{5D}^{on-shell}}{\delta S\left( 0, 0 \right)}e^{-ipx}\\
    &=-\frac{c_g}{2}\sqrt{g_s}e^{-\phi _s\left( z \right)}g_{s}^{zz}\left[ K\left( -\omega,-\boldsymbol{p},z \right) \partial _zK\left( \omega,\boldsymbol{p},z \right) +K\left( \omega,\boldsymbol{p},z \right) \partial _zK\left( -\omega,-\boldsymbol{p},z \right) \right] \mid_{z=0}\\
    &=-\frac{c_ge^{3A_s-\phi _s}}{z^3}f\left( z \right) Re\left[K^*\left( \omega,\boldsymbol{p},z \right) \partial _z K\left( \omega,\boldsymbol{p},z \right)\right]\mid_{z=0},
\end{aligned}
\label{Eq regf}
\end{equation}
where we extracted the value of the fourier transform $S\left( \omega,\boldsymbol{p},z \right)$ of the scalar field 
$S(x,z)$ at the boundary $S_0(\omega,\boldsymbol{p})=S(\omega,\boldsymbol{p},0)$ so that $S\left( \omega,\boldsymbol{p},z \right)$ can be written as the product of its value at the boundary $S_0(\omega,\boldsymbol{p})$ and the propagator in the z-direction $K(\omega,\boldsymbol{p},z)$. Thus $S\left( \omega,\boldsymbol{p},z \right) =S_0\left( \omega,\boldsymbol{p} \right) K\left( \omega,\boldsymbol{p},z \right)$ and $K\left( \omega,\boldsymbol{p},z \right)$ satisfies Eq.(\ref{EoMp}) with boundary conditions Eq.(\ref{Eq bc0 of eomt})($k_0=1$) and Eq.(\ref{Eq bczh of eomt}). In addition, the third equality of Eq.(\ref{Eq regf}) holds because $K^{*}\left( \omega,\boldsymbol{p},z \right)=K\left(- \omega,-\boldsymbol{p},z \right)$.

To solve for $K\left( \omega,\boldsymbol{p},z \right)$, we need to specify the boundary conditions. Near the boundary $z =0$, the expansion of the solution can be obtained from the equation of motion as 

\begin{equation}
    S\left(\omega,\boldsymbol{p},z \right)\rightarrow k_0+\frac{k_0 (\omega ^2-\boldsymbol{p}^2)}{4} z^2 +k_1 z^4-\frac{k_0}{16}(\omega ^2-\boldsymbol{p}^2)[\omega ^2-\boldsymbol{p}^2+6A_s''(0)-2\phi _s''(0)] z^4 \log (z)+...,
\label{Eq bc0 of eomt}
\end{equation}

with \noindent $k_0$ the value of source term. Similarly, near the horizon $z =z_h$, it is not difficult to obtain the expansion as

\begin{multline}
    S\left(\omega,\boldsymbol{p},z \right)\rightarrow \left( z_h-z \right) ^{\mp i\omega \frac{g\left( z_h \right)}{g'\left( z_h \right)}}\left[ 1- \frac{g\left( z_h \right)}{2z_h g' \left( z_h \right) ^2\left( 2\omega g\left( z_h \right) \pm ig' \left( z_h \right) \right)}\right.\\
    \left. \times \left( \left( 2g' \left( z_h \right) ^2\left( \mp ip^2z_h+3\omega -3z_h\omega A_s\prime \left( z_h \right) +z_h\omega \phi _s\prime \left( z_h \right) \right) \pm 2iz_h\omega ^2g\left( z_h \right) g'' \left( z_h \right) \right) -z_h\omega g\prime \left( z_h \right) g'' \left( z_h \right) \right) \left( z_h-z \right)\right.\\
    \left.+O[(z_h-z)^2] \right],
\label{Eq bczh of eomt}
\end{multline}

\noindent with $g(z)=\int_0^{z_h} x^3 e^{-3 A(x)} dx$. It can be seen that there are two solutions for $ S\left(\omega,\boldsymbol{p},z \right)$ in the near horizon expansions, known as the incoming wave condition (negative sign in the exponent) and the outgoing wave condition (positive sign in the exponent), respectively. Choosing the former (latter) corresponds to obtaining the retarded (advanced) correlation function\cite{JHEP09(2002)042}.

When we substitute Eq.(\ref{Eq bc0 of eomt}) into Eq.(\ref{Eq regf}), we find that:

\begin{multline}
    G_R\left( \omega,\boldsymbol{p} \right) = \underset{z\rightarrow 0}{\lim} c_ge^{3A_s-\phi _s}f\left( z \right)\left(\frac{1}{16}(\boldsymbol{p}^2-\omega^2)^2+\frac{-\boldsymbol{p}^2 + \omega^2}{2 z^2} + \frac{4 k_1}{k_0}\right.
    \\
    \left.+ \frac{3}{8} \left( \boldsymbol{p}^2 - \omega^2 \right) A_{s}''(0) - \frac{1}{8} \left( \boldsymbol{p}^2 - \omega^2 \right) \phi_{s}''(0) - \frac{1}{4} \left( \boldsymbol{p}^2 - \omega^2 \right) \ln(z) \left( \boldsymbol{p}^2 - \omega^2 - 6\, A_{s}''(0) + 2\, \phi_{s}''(0) \right)\right)+O(z^2),
\end{multline}

\noindent The terms involving logarithms and $z^{-2}$ can be removed through a regularization procedure; therefore, they are irrelevant for the mass poles and spectral functions. When $\omega$ and $\boldsymbol{p}$ takes specific values such that $k_0=0$, i.e., $S(\omega,\boldsymbol{p},0)=0$, the Green's function diverges, indicating that these values correspond to singularities. Under such boundary conditions, the solution $S(\omega,\boldsymbol{p},z)$ is referred to as a quasi-normal mode, a term that describes the solutions to perturbations in a black hole background, which is consistent with the scenario we are studying, and $\omega$ is its complex frequency.

By setting $\boldsymbol{p}=0$ and $\omega=0$ respectively, we obtain the correlation functions in the temporal and spatial directions. By applying the shooting method to determine $\omega$ and $\left|\boldsymbol{p}\right|$ under the corresponding boundary conditions, we can extract the pole mass, thermal width, and screening mass. If we keep both $\omega$ and $\boldsymbol{p}$ and solve for the complex frequency corresponding to a given $\left|\boldsymbol{p}\right|$, we can obtain the dispersion relation $\omega(\left|\boldsymbol{p}\right|)$. 

Moreover, the retarded Green's function can be used to compute the spectral function:

\begin{equation}
    \rho \left( \omega \right) =-2\,\mathrm{Im}G_R\left( \omega \right).
\label{kubo}
\end{equation}

\noindent If the spectral function exhibits peak structures, the mass and width of the glueball at finite temperature can be extracted. Furthermore, the changes in the spectral function at different temperatures reveal the variations in the properties of glueballs.

In the following, we will present the temperature dependence of the glueball pole mass and thermal width obtained by calculating the QNM frequencies, and simultaneously compute the spectral function for cross-validation. We will then also show the screening mass and dispersion relations at different temperatures.

\subsubsection{Pole mass and thermal width}
\label{sec: Pole mass and thermal width}

\begin{figure}
    \centering
    \includegraphics[width=0.9\linewidth]{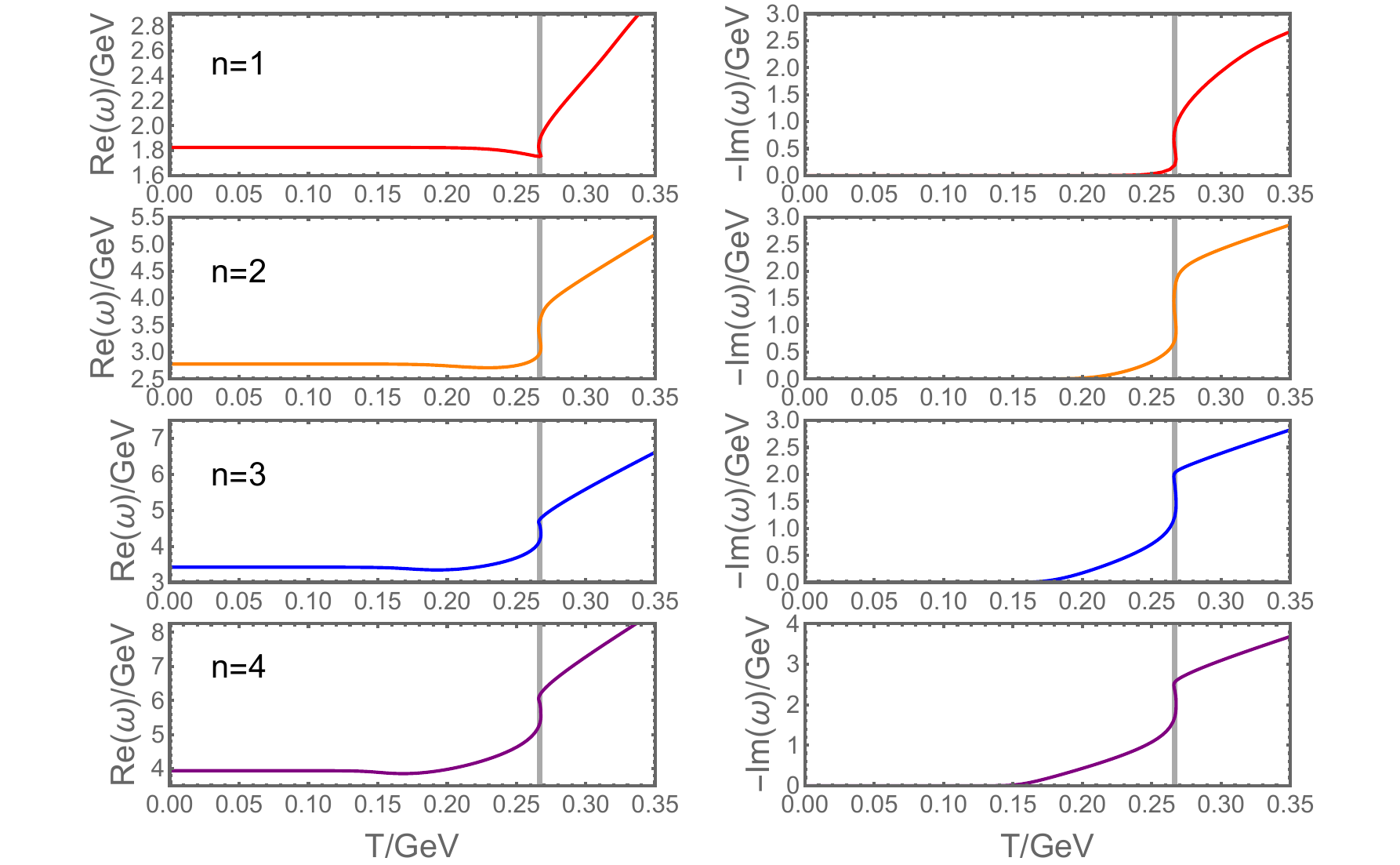}
    \caption{The temperature dependence of the singularities corresponding to the first four glueball bound states at zero temperature are arranged from top to bottom. The left column shows the real parts of the singularities, while the right column shows the imaginary parts.The gray region in the figure indicates where the phase transition occurs.}
    \label{fig:omega-T}
\end{figure}

As mentioned at the beginning of this section, under the condition that boundary condition Eq.(\ref{Eq bczh of eomt}) with $\boldsymbol{p}=0$ is satisfied, the values of $\omega$ that make $S(\omega,0)=0$ correspond to the complex frequencies of the QNMs. The real part of $\omega$ represents the glueball pole mass, while the imaginary part corresponds to the thermal width. Using the shooting method to solve the equation, we can obtain the temperature dependence of several QNM complex frequencies in the case of choice I ($h(z)=e^{-kz^2}$), as shown in Fig.\ref{fig:omega-T}.

At the same temperature, we can obtain many complex frequencies. Among them, a subset has real parts that approach the masses of the glueball bound states at zero temperature, as discussed in the previous subsection, while their imaginary parts approach zero as the temperature tends to zero. This allows us to associate these modes with the mass and thermal width of glueballs at finite temperature. Fig.\ref{fig:omega-T} shows the complex frequencies corresponding to the first four glueball bound states at different temperatures.

These complex frequencies exhibit a common trend as the temperature changes: at low temperatures, the real parts decrease very slowly, making the change almost unnoticeable in the plot—hence, they can be considered approximately constant. Once the temperature increases to a certain value $T_a$ (which we define as the temperature at which the reduction reaches 1\% of the zero-temperature mass), the rate of decrease begins to accelerate, continuing until a point $T_w$, where the real part starts increasing with temperature. The total decrease by this point is about 0.08GeV. After that, as the temperature reaches $T_c$, the real part rises sharply due to the phase transition and then increases linearly with temperature. In contrast, the imaginary parts continuously increase with temperature. However, their growth rate is also very small at low temperatures when the scattering effect is not large enough to produce a large thermal width. Around $T_r$ (which we also define as the temperature at which the width increases by 1 MeV), the rate of increase becomes more significant, followed by a sharp rise near $T_c$, and then a linear increase beyond that point. This indicates the scattering effect from in-medium particles is driving the glueballs to be unstable. We summarize some of the obtained numerical results in Tab.\ref{tab:pole mass}.

\begin{table}
\footnotesize
\renewcommand{\arraystretch}{1.5}
\setlength{\tabcolsep}{1mm}
\begin{tabular}{ccc|ccc|ccc|ccc}
\hline
\multicolumn{3}{c|}{n=1 (m=1.826GeV)} & \multicolumn{3}{c|}{n=2 (m=2.777GeV)} & \multicolumn{3}{c|}{n=3 (m=3.420GeV)} & \multicolumn{3}{c}{n=4 (m=3.941GeV)} \\ \hline
T/GeV & $\omega_R$/GeV & $\omega_I$/GeV & T/GeV & $\omega_R$/GeV & $\omega_I$/GeV & T/GeV & $\omega_R$/GeV & $\omega_I$/GeV & T/GeV & $\omega_R$/GeV & $\omega_I$/GeV \\ \hline
$T_r$=0.218 & 1.819 & 0.001 & $T_r$=0.180 & 2.767 & 0.001 & $T_r$=0.146 & 3.414 & 0.001 & $T_r$=0.138 & 3.924 & 0.001 \\
$T_a$=0.234 & 1.808 & 0.007 & $T_a$=0.195 & 2.751 & 0.009 & $T_a$=0.167 & 3.387 & 0.010 & $T_a$=0.147 & 3.902 & 0.009 \\
\multicolumn{1}{l}{$T_c$=0.265} & 1.755 & 0.147 & $T_w$=0.230 & 2.708 & 0.137 & $T_w$=0.193 & 3.344 & 0.123 & $T_w$=0.169 & 3.854 & 0.131 \\
0.268 & 1.925 & 0.964 & $T_c$=0.265 & 2.913 & 0.627 & $T_c$=0.265 & 4.040 & 1.101 & $T_c$=0.265 & 5.159 & 1.553 \\
0.350 & 2.993 & 2.746 & 0.268 & 3.645 & 1.845 & 0.268 & 4.796 & 2.058 & 0.268 & 6.234 & 2.603 \\
 &  &  & 0.350 & 5.178 & 2.857 & 0.350 & 6.616 & 2.826 & 0.350 & 8.609 & 3.686 \\ \hline
\end{tabular}
\caption{The complex frequencies $\omega_R-i\omega_I $ corresponding to the lowest four scalar glueball bound states at different temperatures. Here, $T_r$ denotes the temperature at which the thermal width reaches 1 MeV, $T_a$ denotes the temperature at which the real part of the complex frequency has decreased by 1\% compared to the corresponding zero-temperature bound state mass, $T_w$ denotes the temperature at which the real part reaches its minimum before $T_c$ which denotes the phase transition temperature.
}
\label{tab:pole mass}
\end{table}

By comparing the behavior of these frequencies, we can observe the following: before $T_c$, both the real and imaginary parts of the frequencies corresponding to higher excited states are consistently larger than those of the lower excited states. Additionally, both $T_a$ and $T_w$ decrease as the excitation number increases. After the phase transition around $T_c$, the ordering of the real parts by magnitude remains unchanged. However, for the imaginary parts, aside from the ground state frequency—which still has the smallest value—the ordering of their widths for the other excited states become uncertain: in our results, the width of the second excited state eventually exceeds that of the third excited state..

\begin{figure}
    \centering
    \includegraphics[width=0.8\linewidth]{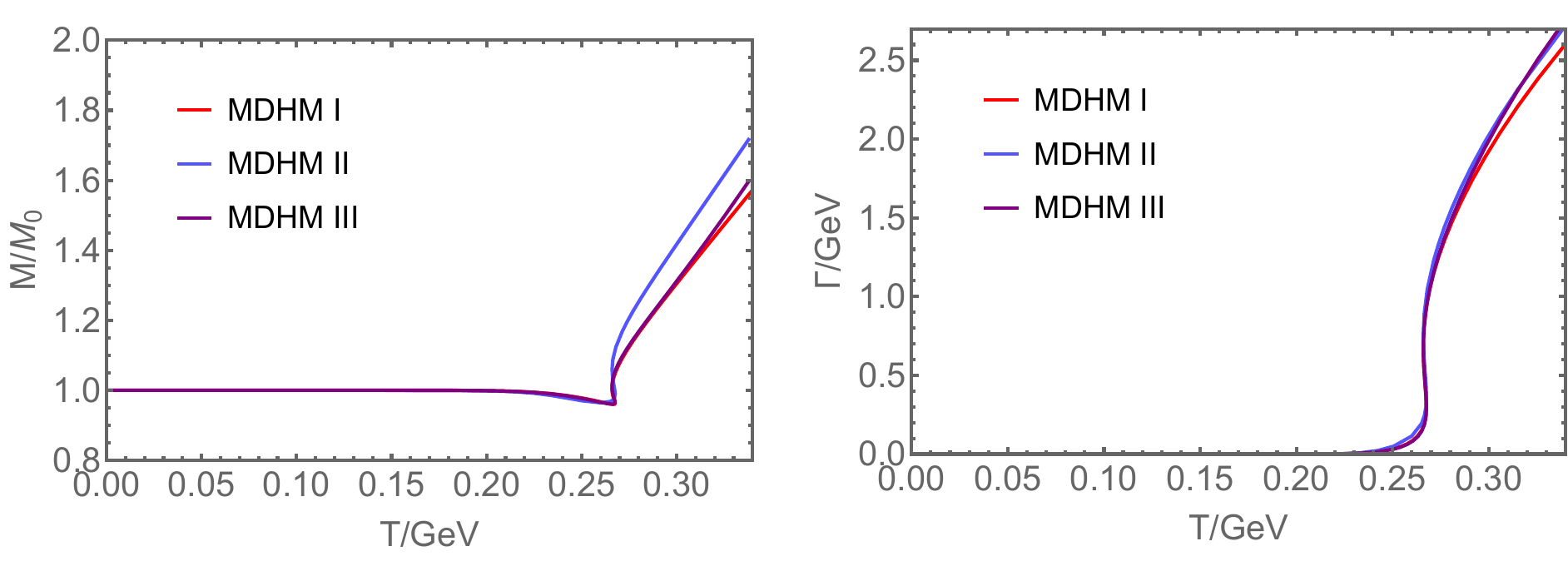}
    \caption{Comparison of the temperature dependence of the glueball ground state mass and width under the three different choices of $h(\phi)$.}
    \label{fig:Compare}
\end{figure}

It is worth noting that although we previously mentioned that there are various possible forms of $h(z)$ that can produce similar glueball spectra at zero temperature. We present in Fig.\ref{fig:Compare} the temperature dependence of the glueball ground state mass and width obtained under the three different choices, which shows that these different choices still yield similar results at finite temperature. Therefore, the results presented above can be considered, to some extent, model-independent. And for the sake of simplicity and its ability to produce a linear spectrum, we will focus primarily on Choice I.

Therefore, we can qualitatively summarize the behavior of the glueball mass and thermal width at finite temperature. The glueball mass remains approximately constant as the temperature initially increases, then begins to decrease slightly, followed by an upward trend. At the critical temperature $T_c$, it undergoes a sharp increase due to the phase transition, and beyond that point, it grows linearly with temperature. Meanwhile, the glueball width continuously increases with rising temperature. However, within the temperature range where the mass remains nearly constant, the increase in width is also minimal. Beyond that, aside from a sharp rise near $T_c$, the growth rate becomes significantly larger and follows a linear trend. 

\begin{figure}
    \centering
    \includegraphics[width=0.8\linewidth]{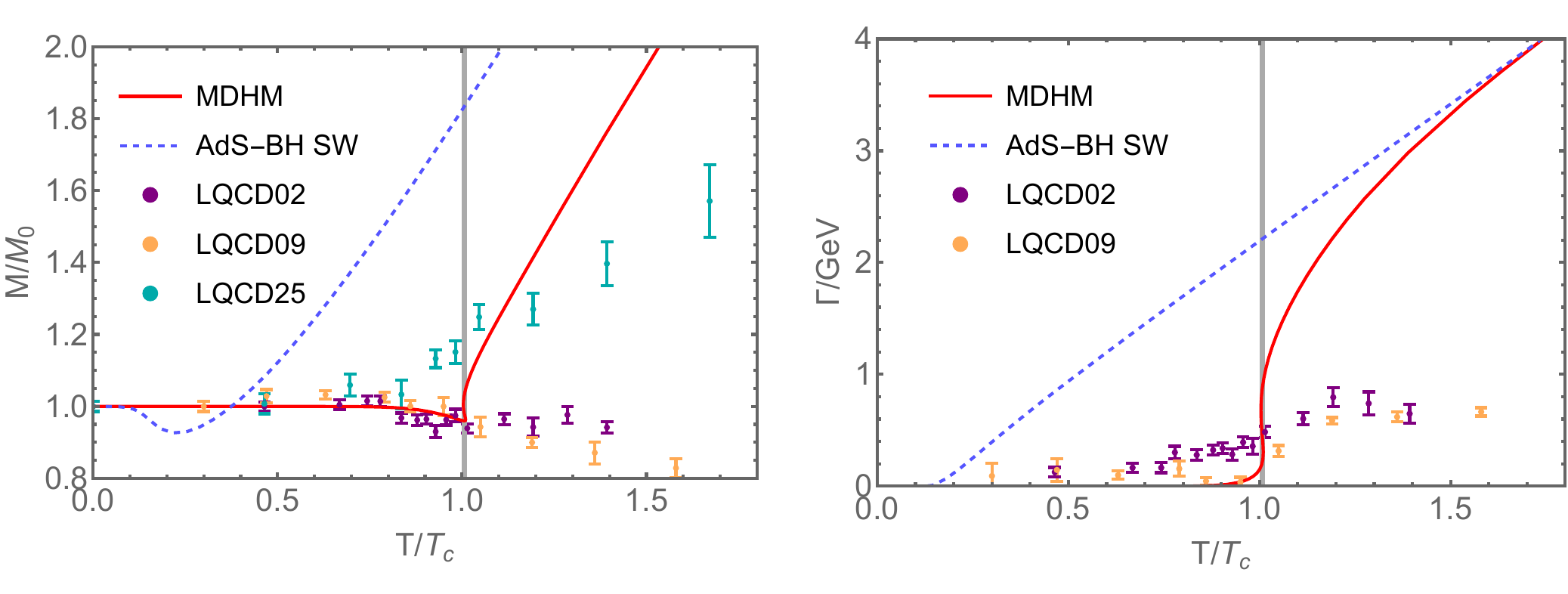}
    \caption{Comparison of the temperature dependence of the scalar glueball ground state mass and width. The red line represents the result obtained in this work using the MDHM model. The blue line shows the result from Ref., where c=0.317GeV. The lattice results shown in purple, orange and cyan are taken from \cite{LQCD02}, \cite{LQCD09} and \cite{LQCD25}, respectively.}
    \label{fig:M&G-T}
\end{figure}

In addition, some existing studies have obtained data on the mass and width of the scalar glueball ground state at finite temperature. Ref.\cite{LQCD02,LQCD09} used lattice QCD combined with Breit-Wigner fitting and concluded that the glueball mass starts to decrease around $T_c$ and continues to decrease at temperatures above 
$T_c$, while the width increases near $T_c$. In contrast, Ref.\cite{LQCD25} additionally considered the effect of the constant term in the correlation function that reflects the deconfinement phase transition, and concluded that the glueball mass remains approximately constant before the phase transition and increases afterward. In addition, \cite{Miranda:TAdSSW09} using holographic methods have investigated the glueball mass and thermal width at finite temperature within the soft-wall model under the AdS-BH(Black Hole) background. These studies found that the mass first decreases and then increases with temperature (with a turning point at around several tens of MeV), while the width increases rapidly throughout. Although different choice of critical temperature $T_c$ and uncertainties in the the ground state mass at zero temperature prevent a fully quantitative comparison, we can still compile the relevant results for qualitative and even semi-quantitative comparison. We summarize the above results in Fig.\ref{fig:M&G-T}. For ease of comparison, we use rescaled coordinates $\frac{M}{M_0}$ and $\frac{T}{T_c}$ in the figure, where $M_0$ represents the mass at the lowest temperature for each dataset. This allows the results at lower temperatures to nearly overlap, making the differences in their temperature dependence more visually apparent.

 It can be seen that before $T_c$, the mass obtained from our model decreases only slightly. Considering the uncertainties in the lattice data, this result is largely consistent with the conclusions from lattice QCD. In contrast, the results from the soft-wall model in the AdS–Black Hole (AdS-BH) background show a clear trend of mass first decreasing and then increasing linearly with temperature, with the turning points occurring at much lower temperatures—leading to significant deviations from the lattice QCD results. After the phase transition at $T_c$, our results show a linear increase in mass with temperature. Although the rate of increase is higher than that in the lattice data [2], it still supports the conclusion that glueball mass begins to rise after $T_c$. As for the width, our results exhibit a similar overall trend to the lattice data, but remain smaller before $T_c$ and then rise rapidly after the transition. Eventually, they reach values much larger than those from lattice QCD, and closely approach the results from the soft-wall model under the AdS-BH background. This indicates that glueballs are more stable before $T_c$, but become rapidly destabilized following the phase transition.

\subsubsection{Spectral function}
\label{subsec: Spectral function}

In this section, we will use the procedure outlined above to calculate the spectral function of the glueball in order to cross-validate the temperature-dependent behavior of the glueball mass and width obtained in the previous section, but we will follow the procedure of the membrane paradigm as explained in Ref.\cite{PhysRevD.79.025023} which is a more straightforward approach.

We first introduce the bulk
response function
\begin{equation}
\zeta(z,\omega)=\frac{\pi(z,\omega)}{\omega S(z,\omega)}, 
\end{equation}
where $\pi(z,\omega)$ is the radial canonical momentum conjugate to the scalar fluctuation $S(z,\omega)$:
\begin{equation}
\pi(z,\omega)=\frac{\delta S_{5D}}{\delta \left( \partial_z S(z,\omega) \right)}=c_g \frac{e^{3A_s-\phi_s}}{z^3} f(z)\partial_z S(z,\omega).
\end{equation}
So the second order differential equation Eq.(\ref{EoMp}) reduced to a first order nonlinear equation:
\begin{equation}
\partial_z\zeta+\frac{z^3 \omega}{e^{3A_s(z)-\phi_s(z)}f(z)c_g}\zeta^2+\frac{e^{3A_s(z)-\phi_s(z)}c_g\omega}{z^3 f(z)}=0.
\end{equation}
Requiring regularity at the horizon, one obtains the following boundary condition:
\begin{equation}
\zeta=i\frac{e^{3A_s(z)-\phi_s(z)}c_g}{z^3}\mid_{z=z_h},
\end{equation}
which is equivalent to the incoming-wave boundary condition.

For the retarded Green function we
have the following relation:
\begin{equation}
G_R\left( \omega \right)=-\frac{c_ge^{3A_s-\phi _s}}{z^3}f\left( z \right) K\left( \omega ,z \right) \partial _zK\left( \omega ,z \right)\mid_{z=0}=\omega \underset{z\rightarrow 0}{\lim} \zeta(\omega,z),
\end{equation}
It is worth noting that, compared to Eq.\eqref{Eq regf}, we have made use of the fact that $K(\omega, z)$ is real at $z = 0$. Then from Eq.\eqref{kubo}, we can obtain the spectral function.

\begin{figure}[ht]
    \centering
    \includegraphics[width=0.9\linewidth]{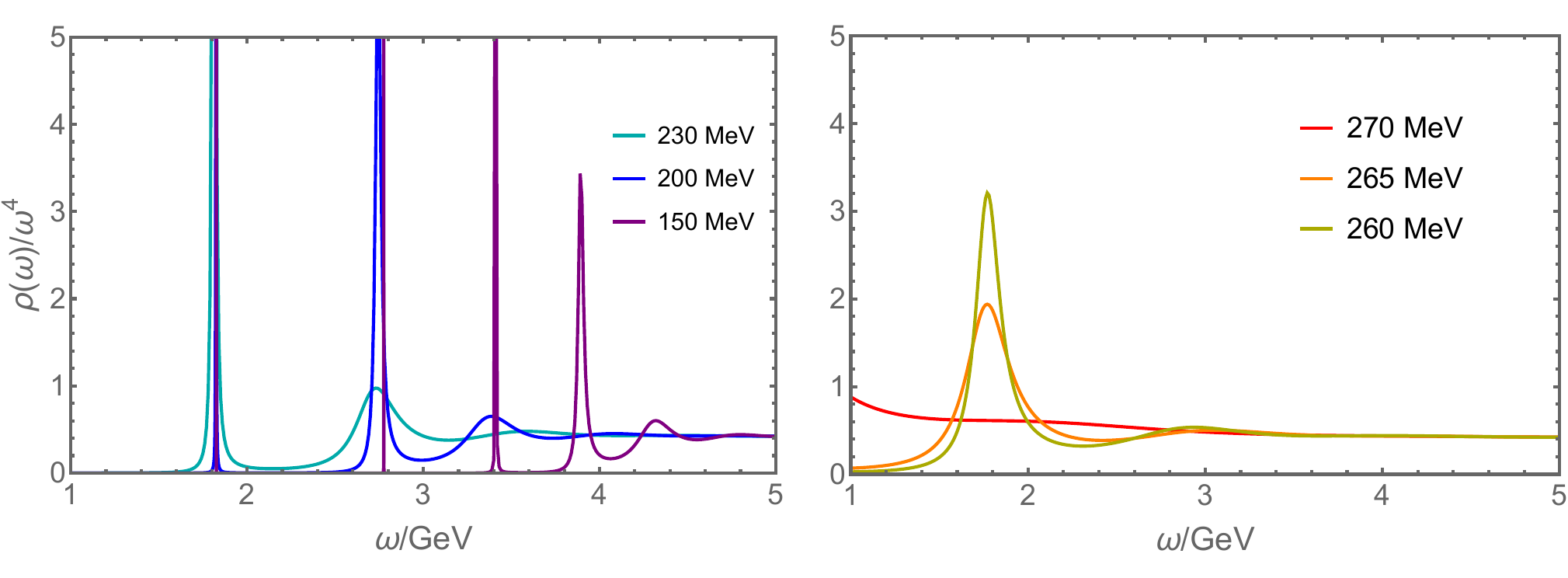}
    \caption{The scaled spectral functions of the scalar glueball at different temperatures is shown. Left panel: from 150 MeV to 230 MeV. Right panel: from 260 MeV to 270 MeV.}
    \label{fig:rho-omega mdhm}
\end{figure}

For the scalar glueball, we have obtained the spectral function shown in Fig.\ref{fig:rho-omega mdhm}. Because for large $\omega$ the spectral function scales as $\omega^4$, we have used the scaled spectral function
\begin{equation}
\tilde{\rho}(\omega)=\frac{\rho(\omega)}{\omega^4}.
\end{equation}

From the graph, we can observe how the spectral function changes with temperature. The spectral function exhibits multiple peak structures, whose central values and widths can be approximately associated with the glueball pole masses and thermal widths. The temperature-dependent behavior of these peaks shows the same patterns as those obtained from the QNM frequency calculations in the previous section: The widths of the peaks broaden as the temperature increases, and at the same temperature, peaks corresponding to lower energies are narrower. Additionally, the positions of the peaks shift leftward at first and then rightward as the temperature rises—this behavior is more obvious for peaks at higher energies.

Moreover, the spectral function provides a more intuitive way to observe the variation in the strength of the glueball two-point correlation function through changes in peak height. Although the QNM frequencies allow us to extract glueball masses and widths at relatively high temperatures, the spectral plots reveal that the peak height decreases as the temperature increases. Once the temperature reaches a certain value, the peak becomes indistinguishable due to the combined effect of its broadening and height reduction—a phenomenon often referred to as "melting" in the literature. This indicates that, at such temperatures, the correlation between gluons becomes so weak that it is hardly discernible from the thermal medium. It can be seen that the higher the excitation level, the lower its melting temperature. After undergoing the phase transition at $T_c$, all states melt due to the drastic increase in their widths.

After obtaining the spectral function, to quantitatively analyze the temperature dependence of the glueball mass and width, it is necessary to extract both by fitting the spectral function using the Breit-Wigner form:

\begin{equation}
    \rho(\omega)=\frac{rm\Gamma\omega^s}{(\omega^2-m^2)^2+m^2\Gamma^2}+t\omega^4,
\end{equation}

\noindent with parameters $m$, $\Gamma$, $r$, $s$, and $t$, where $m$ and $\Gamma$ represent the extracted mass and width, respectively. It is worth noting that although it is not the standard Breit-Wigner formula $\rho(\omega)\sim 1/[(\omega-m)^2+\Gamma^2]$, when the thermal width $\Gamma$ is not too large and the frequency $\omega$ is close to the peak centre, it is equivalent to the standard form and gives a better fitting in our case. In fact, this form has been applied in many other references \cite{Colangelo:2008us, PhysRevD.103.086008}. The last term in the above equation accounts for the influence of the thermal medium (as the spectral function exhibits a $\omega^4$ behavior at large $\omega$). By removing this contribution, the spectral function can be more accurately fitted using the Breit-Wigner formula.

Since we have already obtained the quantitative behavior of the glueball mass and width at finite temperature through the calculation of QNM frequencies, we will not extract these two quantities by fitting the spectral function. However, it should be noted that the results obtained from these two methods may differ, with the discrepancy increasing as the width becomes larger. Typically, the values obtained from the spectral function fitting are greater than those from the QNM method. A detailed discussion has already been provided in Ref.\cite{LQCD02} and \cite{JHEP08(2021)005}.

In summary, the conclusions we obtain from the spectral function are consistent with those derived from the QNM frequency calculations. However, the latter method provides more detailed information, some of which may not be reflected in the spectral function. We discuss this point further in the Appendix.\ref{app:Non-bound-state}. We also want to emphasize that the dissociation of a certain composite particle does not mean the disappear of the corresponding QNM, but the appearance of a large imaginary part (width). Even when the particles dissociate, the corresponding correlations might still be meaningful and can be characterized by the masses and widths.

\subsubsection{Screening mass and dispersion relations}

\begin{figure}
    \centering
    \includegraphics[width=0.9\linewidth]{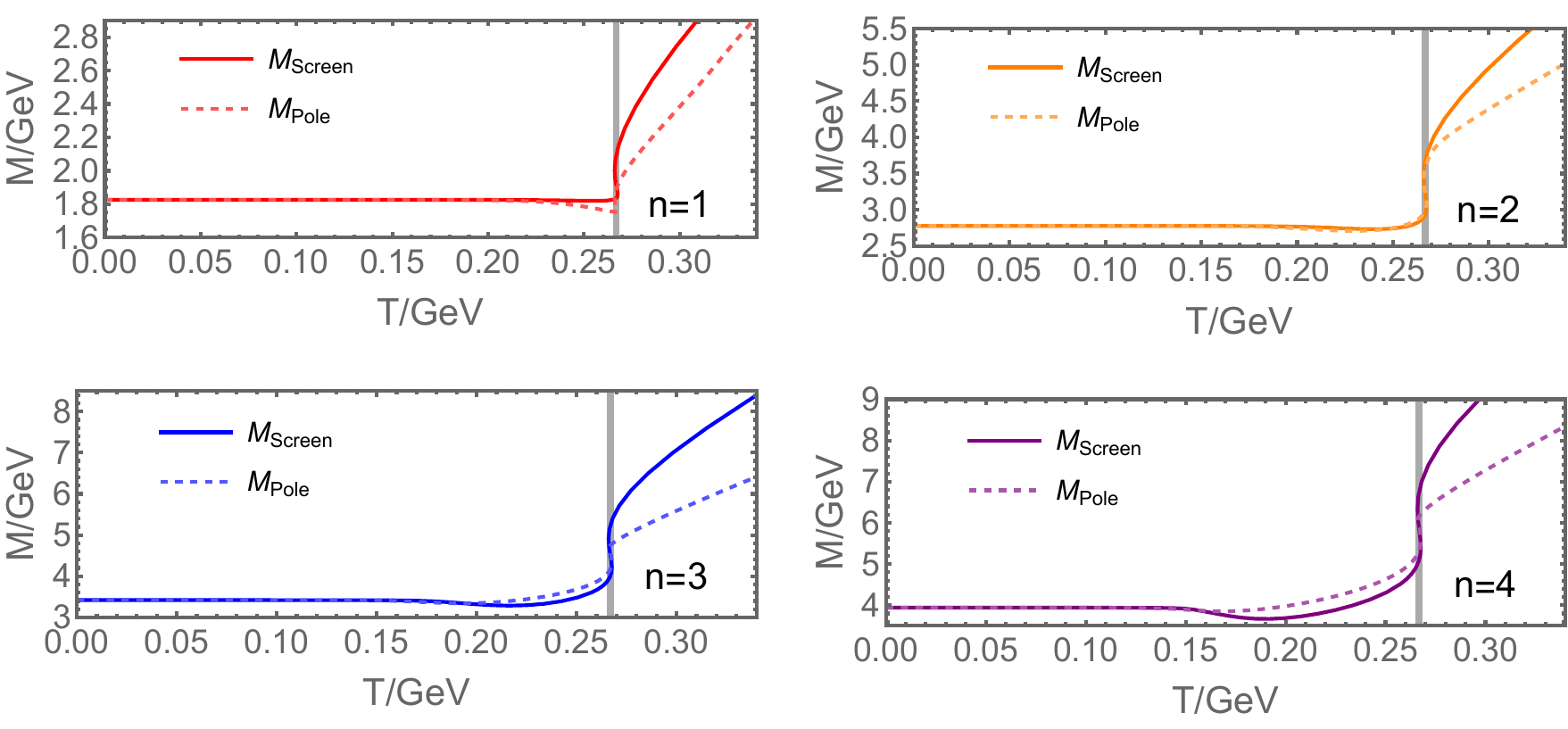}
    \caption{The temperature dependence of the screening masses for each excited state of the scalar glueball, with the corresponding pole mass results plotted as dashed lines for comparison. The gray region in the figure indicates where the phase transition occurs.}
    \label{fig:MDHM Screen Mass}
\end{figure}

If we set $\omega=0$ in Eq.(\ref{Eq bczh of eomt}) and solve for the value of $\left|\boldsymbol{p}\right|$ that satisfies this boundary conditions along with $S(\boldsymbol{p}=i m_{\rm{Screen}},0)=0$, we can then compute the pole of the spatial correlation function and obtain the screening mass $m_{\rm{Screen}}$. The results we obtained are presented in Fig.\ref{fig:MDHM Screen Mass}. For ease of comparison, the pole mass results are plotted together. We also have included part of the screening mass data in Tab.\ref{tab:screen mass}.

From the figure, it can be observed that the screening mass exhibits a temperature dependence similar to that of the pole mass: both masses initially decrease very slightly—so slightly that the change can be considered negligible—then the rate of decrease increases, followed by a reversal where the mass increases with temperature. At the critical temperature $T_c$, a sharp increase occurs, and beyond this point, the mass grows linearly with temperature. Moreover, the higher the excitation level, the greater the extent of both the decrease and the subsequent increase. Therefore, we can adopt the same temperature markers $T_a$ and $T_w$, defined as in the Sec.\ref{sec: Pole mass and thermal width} on pole mass computation, to categorize the temperature ranges where these different behaviors occur. The corresponding values are also included in Tab.\ref{tab:screen mass}.

\begin{table}
\footnotesize
\renewcommand{\arraystretch}{1.5}
\setlength{\tabcolsep}{1.5mm}
\begin{tabular}{ccc|ccc|ccc|ccc}
\hline
\multicolumn{3}{c|}{n=1 (m=1.826GeV)} & \multicolumn{3}{c|}{n=2 (m=2.777GeV)} & \multicolumn{3}{c|}{n=3 (m=3.420GeV)} & \multicolumn{3}{c}{n=4 (m=3.941GeV)} \\ \hline
T/GeV & $m_{s}$/GeV & $\Delta_m$ & T/GeV & $m_{s}$/GeV & $\Delta_m$ & T/GeV & $m_{s}$/GeV & $\Delta_m$ & T/GeV & $m_{s}$/GeV & $\Delta_m$ \\ \hline
$T_w$=0.257 & 1.820 & 2.715\% & $T_a$=0.220 & 2.749 & 1.299\% & $T_a$=0.181 & 3.386 & 0.886\% & $T_a$=0.153 & 3.902 & 0.504\% \\
$T_c$=0.265 & 1.827 & 4.256\% & $T_w$=0.240 & 2.732 & 0.525\% & $T_w$=0.217 & 3.286 & -3.337\% & $T_w$=0.190 & 3.667 & -6.324\% \\
0.268 & 2.152 & 11.776\% & $T_c$=0.265 & 2.869 & -1.667\% & $T_c$=0.265 & 3.864 & -4.543\% & $T_c$=0.265 & 4.911 & -5.009\% \\
0.350 & 3.438 & 14.867\% & 0.268 & 3.804 & 3.598\% & 0.268 & 5.409 & 12.739\% & 0.268 & 7.002 & 12.292\% \\
 &  &  & 0.350 & 6.114 & 18.060\% & 0.350 & 8.699 & 31.490\% & 0.350 & 11.258 & 30.767\% \\ \hline
\end{tabular}
\caption{The screen mass $m_s$ corresponding to the lowest four scalar glueball bound states at different temperatures. Here, $\Delta_m=\frac{m_{s}-m_{pole}}{m_{pole}}$ represents the relative difference between the screening mass and the pole mass with respect to the pole mass, $T_a$ denotes the temperature at which the screen mass has decreased by 1\% compared to the corresponding zero-temperature bound state mass, $T_w$ denotes the temperature at which the screen mass reaches its minimum before $T_c$ which denotes the phase transition temperature.
}
\label{tab:screen mass}
\end{table}

In terms of their relative magnitudes, the screening mass is generally larger than the pole mass throughout most of the temperature range—specifically, in the region where the mass remains approximately constant and beyond $T_c$. However, for all excited states except the ground state, there exists a temperature interval before $T_c$ in which the pole mass exceeds the screening mass.

Meanwhile, if we retain both $\omega$ and $\boldsymbol{p}$ in Eq.(\ref{Eq bczh of eomt}) and solve the equation of motion to obtain a set of $(\omega, \boldsymbol{p})$ data pairs that satisfy $S(\omega, \boldsymbol{p}, 0) = 0$, we can then extract the relation between energy and momentum of the glueball at finite temperature—that is, the dispersion relation. When we plot the dispersion relation curve on the $\boldsymbol{p}^2$-Re($\omega$) plane, the screening mass and the pole mass correspond to the intersections of the curve with the horizontal and vertical axes, respectively. Although the square of the spatial momentum $\vec{p}^2$ is usually taken as real in the dispersion relation, we extend it to the negative half axes in order to connect with the screening pole.

At zero temperature, hadrons follow the relativistic dispersion relation $\omega(\boldsymbol{p}) = \sqrt{\boldsymbol{p}^2 + m^2}$. At low temperatures, however, the dispersion relation is typically modified to the form 

\begin{equation}
\label{REDSFT}
    \text{Re}(\omega(\boldsymbol{p})) = \sqrt{u^2 \boldsymbol{p}^2 + m^2},
\end{equation}

\noindent where $m$ is the pole mass and $u$ is interpreted as a velocity, which is therefore constrained by $u < 1$ \cite{PionLiquid90}. In fact, such a modification is obtained when we neglected the thermal width and the diffusion effect of the corresponding modes. In general, at high temperatures, as we can see below,  due to those effects the dispersion relation will be significantly modified.

\begin{figure}
    \centering
    \includegraphics[width=0.8\linewidth]{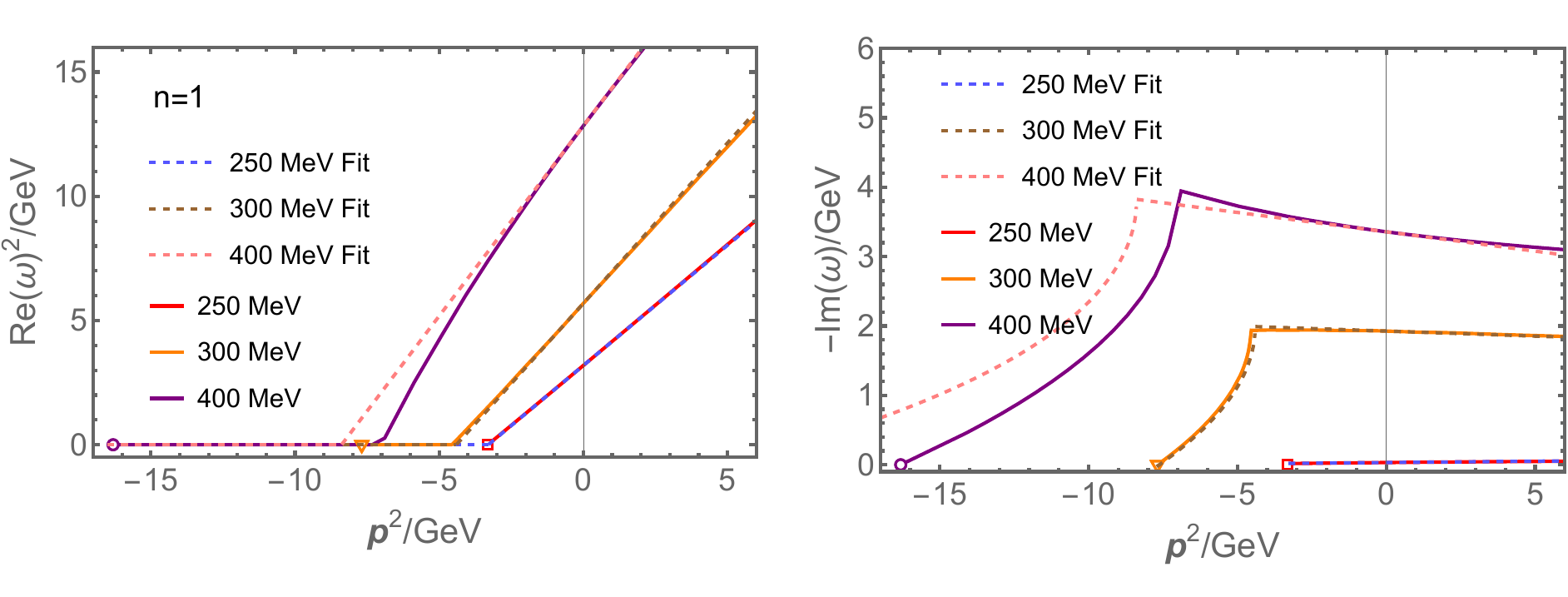}
    \caption{The dispersion relations of the glueball ground state at three different temperatures. Left panel: the relation between the square of the real part of the complex frequency and momentum squared. Right panel: the relation between the imaginary part and momentum squared. We mark the pole positions $(-m_{\text{Screen}}^2, 0)$ at different temperatures in the figure using hollow points. The dashed lines represent the results obtained by fitting the dispersion relations using Eq.\eqref{eq:DSFT}.}
    \label{fig:Dispersion Relation GS}
\end{figure}

We take the ground state as an example and extract the dispersion relations at three different temperatures ($T=250,300,400\rm{MeV}$) in Fig.\ref{fig:Dispersion Relation GS}. To show the behavior of the dispersion relations more explicitly, we plot the square of the real part of the poles and the imaginary part as functions of the momentum square $\vec{p}$. It is found that  $\text{Re}(\omega)^2$ at $T=250 \rm{MeV}$ (the red solid line) and $T=300 \rm{MeV}$ (the orange solid line) almost lie in straight lines with $\vec{p}^2$. Also, in the area of positive $\vec{p}^2$ the imaginary parts are almost linear with $\vec{p}^2$ in the plot region selected. When moving towards the negative momentum square, the square of the real parts reaches zero at a point $\boldsymbol{p}_0^2 > -m_{\text{Screen}}^2$, and remains zero until $\boldsymbol{p}^2 = -m_{\text{Screen}}^2$, where the imaginary part crosses the horizontal axes. Based on those observations, one can summarize the real and imaginary parts in a unique equation as 
\begin{equation}
\label{eq:DSFT}
    \omega(\boldsymbol{p})=\sqrt{u^2 \boldsymbol{p}^2+m^2}-i(\Gamma+D \boldsymbol{p}^2),
\end{equation}
where $m,u,\Gamma, D$ are four parameters. It is easy to understand that $m$ and $\Gamma$ stand for the pole mass and thermal width since they are the real and imaginary parts of $\omega$ when $\vec{p}=0$. At low temperatures, $\Gamma, D$ is very small, since all the diffusion effects in this model comes from the temperature. In this case, the relation will be reduced to Eq. \eqref{REDSFT} and $u$ represents the sound velocity of the corresponding mode. Moreover, since the imaginary part at the real momentum space has the form $\omega\sim -i D \vec{p}^2$, $D$ represents the diffusion constant of the in-medium mode. Of course, all these parameters are temperature dependent quantities. To show this this behavior directly, we also impose a fitting to the numerical data using this formalism, it does give a good description to data, as shown in Fig.\ref{fig:Dispersion Relation GS}. There, the fitting parameters are given as $m=1.786 \text{GeV},\Gamma=0.032 \text{GeV}, u=0.984, D=0.004$ for $T=250\rm{MeV}$ (the blue dashed line) and $m=2.385 \text{GeV},\Gamma=1.927 \text{GeV}, u=1.136, D=-0.015$ for $T=300\rm{MeV}$ (the brown dashed line). From this equation, it is also easy the understand the above numerical results analytically. When $\vec{p}^2>-m^2/u^2$, only contribution to the real part is $\sqrt{u^2\vec{p}^2+m^2}$, and thus its square is linear with $\vec{p}$. Then, if $\vec{p}^2<-m^2/u^2$, $\sqrt{u^2\vec{p}^2+m^2}$ becomes pure imaginary. Thus, the real part disappears and the imaginary part becomes the curve of a square-root type. Here, due to the effect of thermal broadening and diffusion, the screening mass is no longer $m_{\text{screen}}=m/u$. Instead, it should be solved from the vanishing of the imaginary part, i.e. $\sqrt{-u^2 m_{\text{Screen}}^2+m^2}-i(\Gamma-D m_{\text{Screen}}^2)=0$. From this equation one gets a more general relation between the two mass parameters with the corrections from thermal effects, and it reads
\begin{equation}
    m_{\text{Screen}}^2=\frac{2(m^2+\Gamma^2)}{u^2+2D\Gamma+\sqrt{u^4-4D^2m^2+4D\Gamma u^2}}.
\end{equation}
At low temperatures when $D$ and $\Gamma$ are sufficiently small, it reduces to $m_{\text{Screen}}=m/u$. 

Finally, for $T=400 \rm{MeV}$ (the purple solid lines in Fig.\ref{fig:Dispersion Relation GS}), we find that both the real and imaginary part exihibit certain non-linear behavior with $\vec{p^2}$, and Eq. \eqref{eq:DSFT} (with $m=3.583 \text{GeV},\Gamma=3.357 \text{GeV}, u=1.237, D=-0.056$, the pink dashed lines) can only partially describe the numerical data. Of course, by considering more powers of $\vec{p}^2$, it is possible to describe those high temperature poles. However, since at such high temperatures, the particles might already dissociate, we will not go deeper in this direction.

Finally, we note that in our calculation, we find that when $\vec{p}^2<-m^2/u^2$, there is another branch of solution corresponding to choosing an opposite sign in the square root. This indicate that the full progator at finite temperature is proportional to 
\begin{equation}
    G(\omega,\vec{p})\sim\frac{1}{[\omega+i(\Gamma+D \vec{p}^2)]^2-u^2\vec{p}^2-m^2},
\end{equation}
which includes the propagation and diffusion effect together (we do not try to get information of the decay constants in the numerator). Though from the thermal field theory, it is also possible to obtain such a correction from the imaginary part of the self-energy, it gives a more convenient way to get the thermal coefficients from the holographic framework. It is interesting to study the effects from those corrections on the thermodynamics. We will leave it to the future.

\section{Discussion and Conclusion}
\label{sec:Discussion and Conclusion}

In this work, based on a machine learning holographic QCD model, a systematical framework has been constructed to investigate the properties of the scalar glueballs continuously from zero temperature to finite temperature. The thermal scalar glueballs are considered as excitations on gravitational backgrounds. These backgrounds are obtained by assuming a specific form of the warp factor A(z) in the gravity-dilaton system and have been carefully tested in describing the thermodynamic quantities of the pure gluon system \cite{PhysRevD.109.L051902}. The coupling function $h(\Phi_s)$ between the glueball sector and the background is fixed by the vacuum spectrum of the glueballs, and it is found to provide a good description of the spectrum. Subsequently, the pole mass, screening mass, thermal widths, and dispersion relations—quantities that characterize the correlations of glueballs—are computed and subjected to a detailed analysis by using both the quasi-normal frequencies and the spectral functions. 

For the temporal components, we find that the pole mass remains approximately constant at first as the temperature increases, then decreases slightly, and eventually increases again—with a sharp rise at the critical temperature $T_c$, after which it grows linearly. Meanwhile, the width consistently increases with temperature, except for a rapid growth during the phase transition. In particular, the temperature dependence of the ground state shows good agreement with lattice results below $T_c$, and supports the recently proposed linear growth behavior above $T_c$. By employing three distinct forms of the coupling function, we observe that this qualitative behavior is, to some extent, independent of the specific form of $h(\Phi_s)$. This finding may lend support to earlier lattice results \cite{Ishii:2001zq,Ishii:2002ww,Meng:2009hh} and indicates that the contribution of constant terms at low temperatures \cite{Arikawa:2025kjx} warrants further investigation. For the spatial component, we found that the screening mass exhibits a similar trend to the pole mass. Moreover, except for a narrow temperature range before $T_c$ for the excited glueball states, the screening mass is always larger than the pole mass. For the dispersion relation, we find that at relatively low temperatures, it can be well described by the formula
$\omega(\vec{p}^2) = \sqrt{u^2 \vec{p}^2 + m^2} - i(\Gamma + D \vec{p}^2)$, which indicates that the hot meson correlators might have the form of $G(\omega, \vec{p})\sim\frac{1}{[\omega+i(\Gamma+D \vec{p}^2)]^2-u^2\vec{p}^2-m^2}$. As the temperature increases, however, the nonlinear features of the dispersion relation become more prominent.

Finally, as mentioned in the introduction, previous holographic models have certain limitations in their range of applicability. However, by modifying the model in the manner described above, we appear to have the potential to overcome these limitations. Our model not only yields black hole solutions valid across the entire temperature range from zero to infinity and produces an equation of state that agrees well with lattice results, but also reproduces glueball spectra and their temperature dependence in close agreement with lattice QCD. By using the couping function of the form $e^{-kz^2}$, our model construction becomes somewhat similar to the soft-wall model in the AdS-BH background, while we adopt a more sophisticated asymptotically AdS metric for the soft-wall model, which leads to significant improvements—since the original soft-wall model fails to properly describe the glueball spectrum and its finite-temperature behavior deviates substantially from lattice results—our modifications still yield notably better agreement.

\appendix
\section{Non-bound-state poles in QNM frequency and spectral functions method}
\label{app:Non-bound-state}

In the main text, we have presented the calculation of the glueball pole mass and thermal width at finite temperature using both the QNM frequency method and the spectral function method. We also highlighted the characteristics of each approach: the former allows for more accurate computations and is applicable over a wider temperature range, while the latter, though becoming increasingly difficult to apply as the peak width grows, offers a more intuitive representation  since it is in the real frequency space.

However, during the calculation process, we discovered another feature of the QNM method. In the main text, we focused on the poles whose real parts approach the zero-temperature glueball masses and whose imaginary parts approach zero as temperature decreases. In fact, we also identified other types of poles whose real and imaginary parts both tend toward zero at zero temperature, and whose behavior with respect to temperature differs from those that correspond to glueball masses. We will refer to them as non-bound-state poles for now. Nevertheless, the effects of them may not necessarily be reflected in the spectral function.

\begin{figure}
    \centering
    \includegraphics[width=0.8\linewidth]{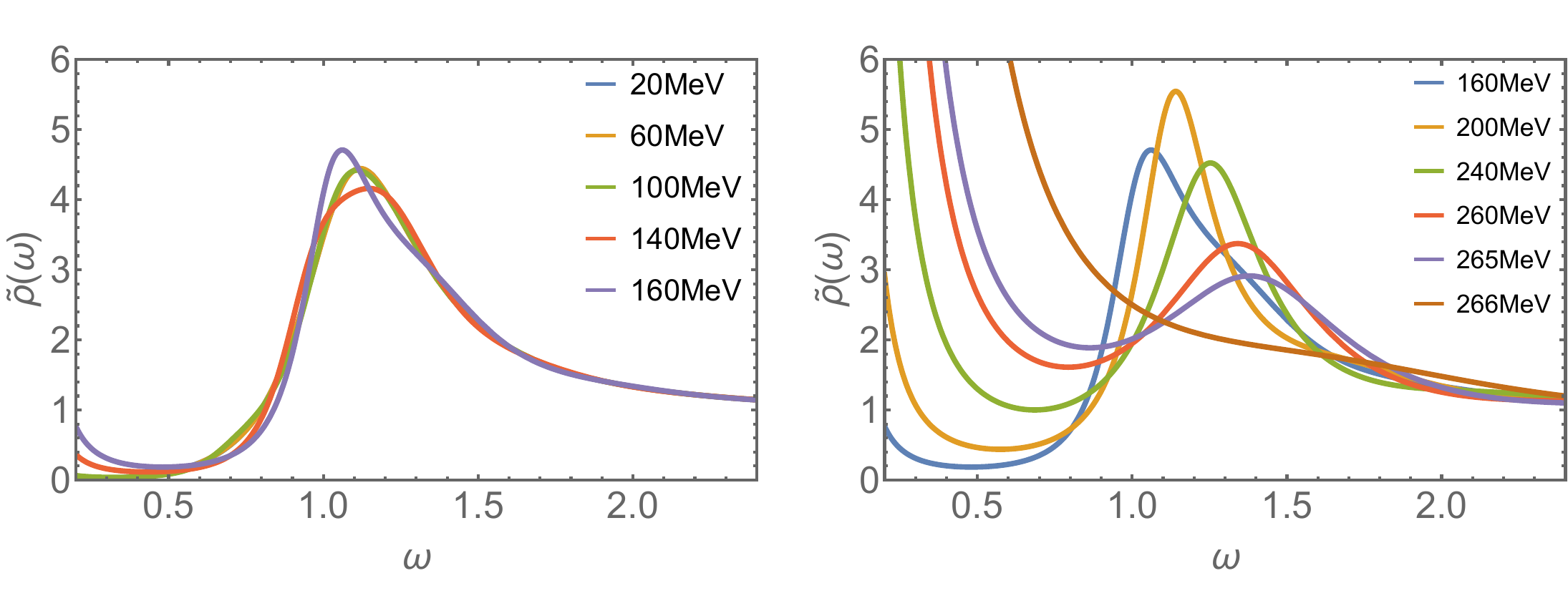}
    \caption{Spectral functions at different temperatures obtained with $A(z)=d\, ln(bz^4+1)$ in the dynamical holographic model.}
    \label{fig:rho-omega s}
\end{figure}

\begin{figure}
    \centering
    \includegraphics[width=0.9\linewidth]{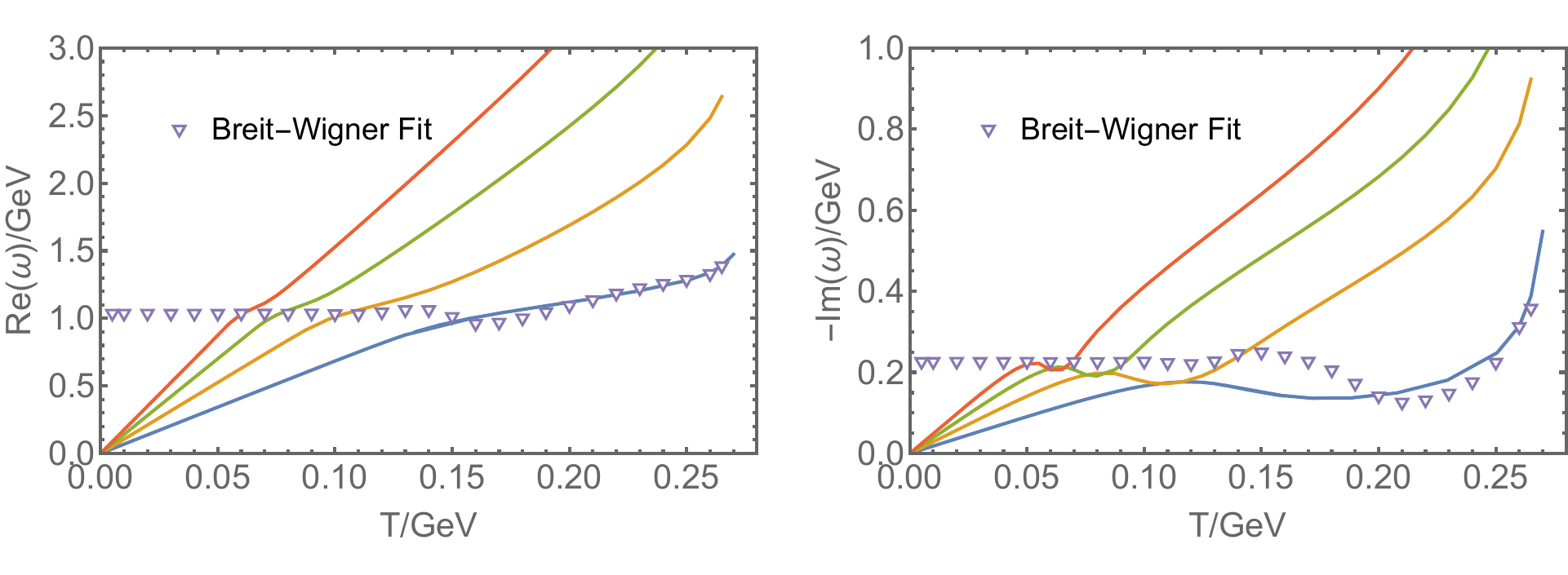}
    \caption{Temperature dependence of four pole positions in the correlation function obtained via the QNM complex frequency method(Colored solid lines), compared with the results extracted from the spectral function using Breit-Wigner fitting(purple inverted triangles). In addition, many other poles with similar behavior have been omitted.}
    \label{fig:QNM-Spec DHM}
\end{figure}

Let us first give an example that clearly illustrates this feature. When using the model presented in the main text with $h(\Phi)=e^{-\Phi_s}$, i.e., the unmodified dynamical holographic model, we obtain the spectral function shown in Fig.\ref{fig:rho-omega s}. 
From this figure, we can clearly observe a peak structure in the spectral function, which shows little variation at low temperatures but both the peak position and its width increase with rising temperature at higher temperatures. At low temperatures, the center of this peak is slightly above 1.0 GeV, which might suggest that it corresponds to the glueball ground state. However, as we showed in Sec.\ref{subsec Glueball at zero Temperature}, this system cannot produce bound states at zero temperature, so such a correspondence lacks a reasonable foundation. If we instead determine the positions of the poles by calculating the QNM complex frequencies, we obtain the results in Fig.\ref{fig:QNM-Spec DHM}.
It is noticed that there are other QNMs, but we only show the relevant ones.

In this figure, we show only a subset of the pole evolutions; in addition to these, there exist many other poles with different values but similar behavior. Using this method, we observe that all the poles approach zero as the temperature decreases, but there is no pole that tends toward a constant value. When we compare these results with those obtained from fitting the spectral function using the Breit-Wigner method (as described in detail in Sec.\ref{subsec: Spectral function}), we find that at higher temperatures, the fit closely matches the lowest QNM frequency, while at lower temperatures, a noticeable deviation occurs. We can conclude that the peak structure of the spectral function at low temperatures in this model results from the combined contributions of multiple poles, which causes it to remain nearly unchanged over a considerable temperature range. As the temperature increases, the spacing between these poles grows, and the peak gradually becomes dominated by the lowest-lying pole. During the transition between these two regimes, a decrease in both the peak position and width appears—this is also evident in the unusual shape of the spectral function at 160 MeV shown in Fig.\ref{fig:rho-omega s}.

\begin{figure}
    \centering
    \includegraphics[width=0.9\linewidth]{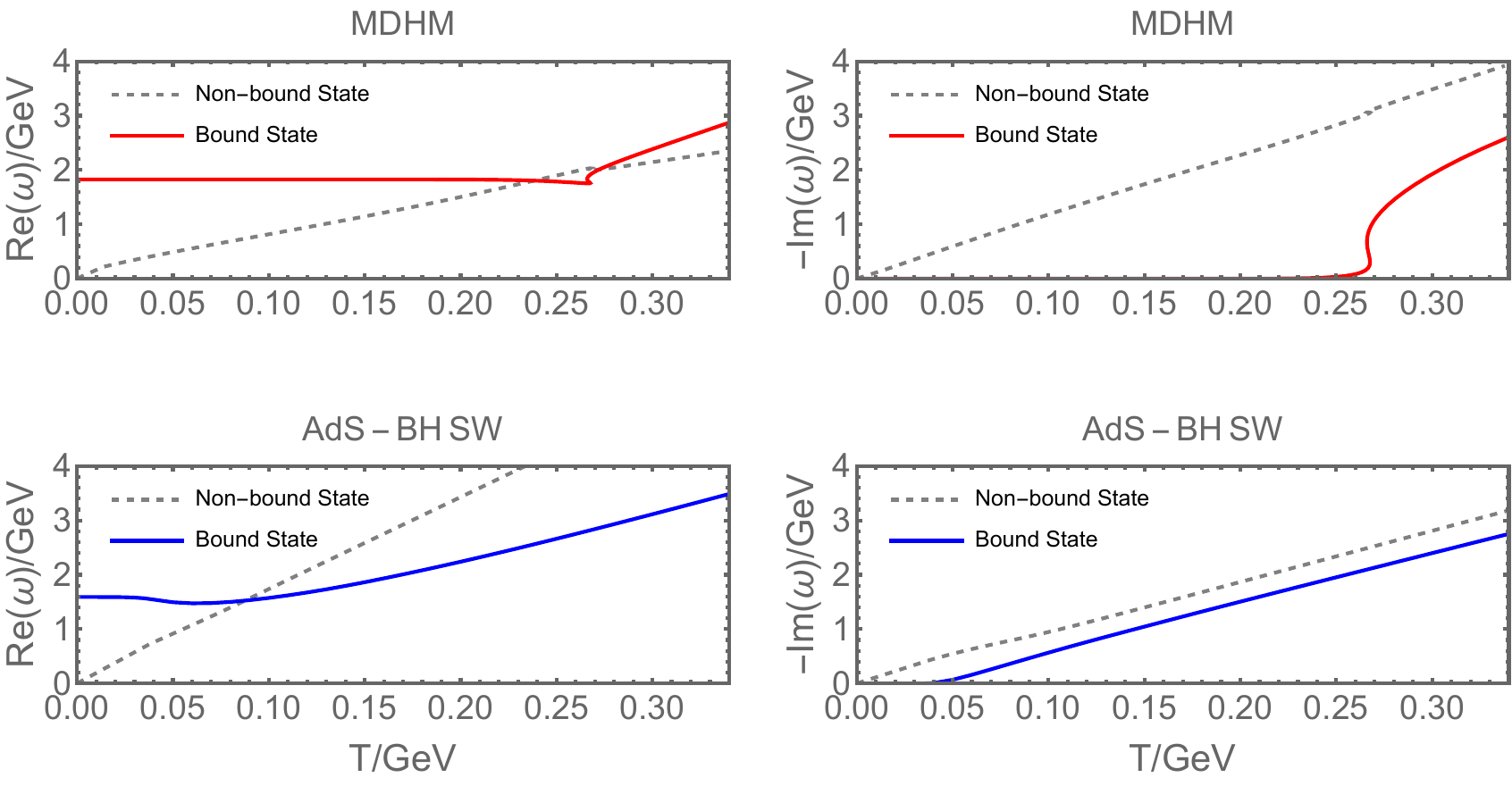}
    \caption{Comparison of the temperature dependence between a non-bound-state pole and the pole corresponding to the ground state in the improved dynamical holographic model (MDHM) and the soft-wall model under the AdS-BH background (AdS-BH SW).In each model, the non-bound-state poles are indicated by gray dashed lines, while the bound-state poles are represented by solid lines.}
    \label{fig:Non-bound state with confine}
\end{figure}

Returning to the modified dynamical holographic model used in the main text and the soft-wall model under the AdS-BH background, we can likewise observe the existence of such poles that do not correspond to zero-temperature bound states. There are likewise many such poles at a given temperature. We select one of them and compare its temperature-dependent behavior with that of the pole corresponding to the ground state. The results are shown in the fig.\ref{fig:Non-bound state with confine}. These non-bound-state poles exhibit relatively consistent behavior: both their real and imaginary parts increase approximately linearly with temperature. However, the growth slopes differ among different poles. Additionally, since the metric in the MDHM model encodes information about the phase transition, the slopes also change before and after the transition.

However, unlike the previous example, these poles do not appear to have any noticeable effect on the spectral function in this case. This is because, in the two models capable of producing bound states, the imaginary parts of the non-bound-state poles grow much faster with temperature—especially at low temperatures—than those of the bound-state poles. As a result, a "melting" effect occurs at very low temperatures: the width at which bound states melt is often already reached by the non-bound-state poles at around 10 MeV. Therefore, within the temperature range of interest, the influence of non-bound-state poles on the spectral function becomes negligible. In contrast, in the model from the first example, the growth rate of the imaginary parts of non-bound-state poles is significantly slower, allowing visible peak structures to appear even before $T_{c}$. It remains unclear whether this peculiar property is related to the model’s inability to produce bound states.

Physically, the states corresponding to these non-bound-state poles can be interpreted as certain excitations of the thermal medium. At finite temperature, these states typically possess relatively large thermal widths, leading to rapid decay; as a result, they usually produce observable effects only at very low temperatures. However, at higher temperatures, as the widths of bound states increase rapidly, it is possible for some of these excitations—with widths smaller than those of the bound states—to become the dominant contributors to the spectral function (albeit very weakly at that stage).

Moreover, the appearance of these non-bound-state poles highlights another advantage of the QNM frequency method: its ability to comprehensively identify all the poles of the two-point correlation function. This is something that cannot be achieved through the spectral function method alone. Relying solely on the latter may even lead to misleading interpretations, as illustrated in the first example where non-target states dominate the spectral signal. Therefore, in practical studies, a combination of multiple methods is essential to obtain more robust and reliable conclusions.

\begin{acknowledgments}
 This work is supported in part by the National Natural Science Foundation of China (NSFC) Grant Nos. 12235016, 12221005, and 12275108, and the Fundamental Research Funds for the Central Universities. 
\end{acknowledgments}

\bibliographystyle{unsrt}
\bibliography{Glueball}

\end{document}